\newcommand*{\ditto}{\raisebox{-0.5ex}{\ttfamily"}}

\documentclass[manuscript]{aastex6}
\usepackage{amsbsy}
\usepackage{amsfonts}
\usepackage{amssymb}
\usepackage{graphicx}
\usepackage{graphics}
\usepackage{bm}

\begin{document}
\title{Electron-Impact Multiple Ionization Cross Sections for Atoms and Ions of Helium through Zinc}
\author{M. Hahn\altaffilmark{1}, A. M\"uller\altaffilmark{2}, and D. W. Savin\altaffilmark{1}}
\altaffiltext{1}{Columbia Astrophysics Laboratory, Columbia University, 550 West 120th Street, New York, NY 10027 USA} \email{mhahn@astro.columbia.edu}
\altaffiltext{2}{Institut f\"{u}r Atom- und Molek\"{u}lphysik, Justus-Liebig-Universit\"{a}t Giessen, Leihgesterner Weg 217, 35392 Giessen, Germany}

\date{\today}
\begin{abstract}

We have compiled a set of electron-impact multiple ionization (EIMI) cross sections for astrophysically relevant ions. EIMI can have a significant effect on the ionization balance of non-equilibrium plasmas. For example, it can be important if there is a rapid change in the electron temperature or if there is a non-thermal electron energy distribution, such as a kappa distribution. Cross sections for EIMI are needed in order to account for these processes in plasma modeling and for spectroscopic interpretation. Here, we describe our comparison of proposed semiempirical formulae to the available experimental EIMI cross section data. Based on this comparison, we have interpolated and extrapolated fitting parameters to systems that have not yet been measured. A tabulation of the fit parameters is provided for 3466 EIMI cross sections. We also highlight some outstanding issues that remain to be resolved.  
		
\end{abstract}

	
\maketitle
	
\section{Introduction}\label{sec:intro}

Collisionally ionized plasmas are those formed by electron-impact ionization (EII). Such plasmas are common in astrophysical sources, such as stars, supernova remnants, galaxies, and galaxy clusters. Modeling the emission from these objects requires knowing the charge state distribution (CSD) within the plasma, which is set by the ionization and recombination rates \citep{Bryans:ApJS:2006, Bryans:ApJ:2009}. 

Electron-impact multiple ionization (EIMI) is the EII process in which a single electron-ion collision results in the ejection of multiple electrons. In most previous work EIMI has been ignored. For plasmas near thermal equilibrium, the multiple ionization rates for a given charge state are insignificant at the temperatures where that charge state is most abundant. However, EIMI can be important in dynamic systems where ions are suddenly exposed to higher electron temperatures \citep{Muller:PhysLett:1986, Hahn:ApJ:2015}. For this reason, EIMI may be important for studies of solar flares \citep{Reale:ApJ:2008, Bradshaw:ApJS:2011}, nanoflare coronal heating \citep{Hahn:ApJ:2015}, supernova remnants \citep{Patnaude:ApJ:2009}, and merging galaxy clusters \citep{Akahori:PASJ:2010}. EIMI can also have a significant effect on the CSD for plasmas with a non-thermal electron energy distribution. For such plasmas there is a substantial population of electrons in the high energy tail of the distribution that lie above the EIMI threshold \citep{Hahn:ApJ:2015b}. Thus, EIMI is relevant to the modeling of astrophysical systems where such non-thermal distributions are present.

Due in part to a lack of data, EIMI has been largely ignored, despite its relevance to astrophysical models. One reason for the lack of EIMI data is that theoretical calculations for EIMI are very challenging. There are at least four particles whose interactions must be accounted for: the ion, incident electron, and two or more ejected electrons \citep{Berakdar:PhysLett:1996, Muller:NIMB:2005, Gotz:JPhysB:2006}. This means that there are at least three electrons in the continuum. Because of these difficulties, theoretical calculations have mainly been performed for simple systems \citep[e.g.,][]{Defrance:JPhysB:2000, Pindzola:JPhysB:2009, Pindzola:JPhysB:2010, Pindzola:JPhysB:2011}, though recent work is attempting to extend these calculations to more complex systems \citep{Pindzola:JPhysB:2017}.  

Most EIMI cross sections are based on experimental measurements. For astrophysical applications EIMI data are needed for nearly all the charge states of all the elements from He--Zn. However, it is not possible for experiments to measure all of these data. In order to estimate cross sections for unmeasured systems, various semiempirical formulae have been proposed. 

Here, we have produced a database of EIMI cross sections for astrophysics. In order to generate this database, we compared measurements of EIMI cross sections to the predictions using the proposed semiempirical formulae. Table~\ref{table:refs} gives references for all of the experimental data sources that we consulted. Based on this comparison, we selected the semiempirical scheme that worked best for a given isoelectronic sequence and EIMI process (e.g., double, triple, etc...) and then used that scheme to generate EIMI cross sections for the unmeasured systems. The procedures are described below in Section~\ref{sec:cross} and their application to each isoelectronic sequence is discussed in Section~\ref{sec:app}. The main results of this work are the tabulated cross section parameters, which are available as online data, a selection of which is illustrated in Table~\ref{table:cross}. Section~\ref{sec:conclusions} concludes with a brief summary of the state of the EIMI data and avenues for future improvements.

\section{Electron-Impact Multiple-Ionization Cross Sections}\label{sec:cross}

EIMI can proceed via several different processes \citep{Muller:Book:2008}. These processes can be broken into the categories of direct ionization (DI) and indirect processes such as excitation-multiple-autoionization (EMA), ionization-autoionization (IA), and multiple-ionization-multiple-autoionization (MIMA). This classification is oversimplified, as even direct multiple ionization can be broken into a sequence of processes. For example, direct ionization has often been considered to proceed via two-step mechanisms, called TS1 and TS2 \citep{Gryzinksi:PR:1965}. In TS1 the incident electron collides with and removes one electron, and then that previously bound electron collides with and removes a third electron. In TS2 the incident electron collides with and removes both bound electrons sequentially. Recently, \citet{Jonauskas:PRA:2014} have extended this picture to include further processe, such as ionization-excitation-ionization and excitation-ionization-ionization. It is difficult to distinguish these processes in total cross section measurements, although differential cross section experiments can provide some information \citep[e.g.,][]{Lahmam:JPhysB:2010}. 

We have found that nearly all of the existing EIMI cross section measurements can be adequately represented as a sum of direct ionization and IA cross sections. The relative contribution of these processes depends on the isoelectronic sequence and the charge state. For example, double ionization of ions in low charge states is dominated by direct ionization, while for more highly charged ions, the IA process dominates \citep[e.g.,][]{Muller:PRL:1980,Cherkani:PhysScr:2001,Hahn:ApJ:2011}. Although other processes such as EMA, have been identified in some measurements \citep{Muller:PRL:1988}, incorporating these effects into our representations of the cross sections would be cumbersome as there is insufficient data upon which to base an interpolation. Fortunately, the resulting cross section errors are usually of little impact.

\subsection{Fitting Formulae}\label{subsec:form}

In order to represent the EIMI cross sections, we use various semiempirical formulae. Direct EIMI can be described by \citep{Shevelko:PhysScr:1995, Shevelko:JPhysB:1995, Belenger:JPhysB:1997}, 
\begin{equation}
\sigma_{\mathrm{D}}=f \frac{p_{0} p_{1}^{p_{2}}}{\left(E_{\mathrm{th}}/E_{\mathrm{Ryd}}\right)^2}\left(\frac{u+1}{u}\right)^{p_{3}}\frac{\ln{(u)}}{u} 
\times 10^{-18} \,\mathrm{cm^{2}},
\label{eq:shevtar}
\end{equation}
where $u=E/E_{\mathrm{th}}$ is the incident electron energy $E$, in eV, normalized by the direct multiple-ionization threshold $E_{\mathrm{th}}$ and $E_{\mathrm{Ryd}}=13.606$~eV. The parameters $p_{0}$ and $p_{2}$ depend on the number of electrons being removed and have been tabulated by \citet{Shevelko:JPhysB:1995} and \citet{Belenger:JPhysB:1997}. The parameter $p_{1}$ is the number of electrons in the target ion and $p_{3}=1.0$ for neutral targets or $0.75$ for ionic targets. \citet{Shevelko:PhysScr:1995, Shevelko:JPhysB:1995} and \citet{Belenger:JPhysB:1997} obtained these fitting parameters by comparing with relatively heavy targets, such as Kr$^{0+}$, Xe$^{0+}$, and Rb$^{1+}$. For lighter systems, we have found that this formula often overestimates the cross section, so we have introduced an additional scaling factor $f$ in order to obtain a better fit to the experimental data. An advantage of this formula is that it can represent ejection of between two and ten electrons, whereas other formulae are best suited only for double ionization.

Direct double ionization of light ions is often more accurately described by \citep{Shevelko:JPhysB:2005},
\begin{equation}
\sigma_{\mathrm{D}}=\left[1-\mathrm{e}^{-3\left(u-1\right)}\right] \left\{ \frac{p_0}{E_{\mathrm{th}}^3}\left[\frac{u-1}{\left(u+0.5\right)^2}\right] \right\}
\times10^{-13} \, \mathrm{cm^{2}}.
\label{eq:shev5a}
\end{equation}
Here, $E_{\mathrm{th}}$ is the threshold for direct EIMI and $p_{0}$ and $p_{1}$ are fitting parameters. 
\citet{Shevelko:JPhysB:2005} give parameters for $p_{0}$ that vary depending on the initial isoelectronic sequence and are valid for He-like through Ne-like ions as well as for Ar$^{0+}$ -- Ar$^{7+}$. As described below, we have found that, with a few slight adjustments, their argon fitting parameters also describe reasonably well the rest of the ions in the corresponding isoelectronic sequences. For near neutrals, we have found that the $p_{0}$ values given by \citet{Shevelko:JPhysB:2005} systematically underestimate the direct double ionization cross sections and we provide new $p_{0}$ values for those cases where there are sufficient data, a point that will be further illustrated in Section~\ref{sec:app}. 

\citet{Shevelko:JPhysB:2005} also gave a semiempirical formula for IA causing double ionization of light ions, which is given by
\begin{equation}
\sigma_{\mathrm{IA}}=f_{\mathrm{BR}} \frac{p_{0}}{E_{\mathrm{th}}^2}\frac{u-1}{u \left(u+p_{1}\right)} 
\times10^{-13} \, \mathrm{cm^{2}},
\label{eq:shev5b}
\end{equation}
where $E_{\mathrm{th}}$ is the IA threshold, i.e., the threshold for single ionization of a core electron forming an intermediate state that can autoionize. The parameters $p_{0}$ and $p_{1}$ depend on the isoelectronic sequence of the initial ion configuration. Here, we introduce the quantity $f_{\mathrm{BR}}$, which is the branching ratio for autoionization of the intermediate state that is missing an inner-shell electron. 

We generally take the IA branching ratios from \citet{Kaastra:AAS:1993}, although other calculations of branching ratios exist \citep[e.g.,][]{Bautista:AA:2003, Gorczyca:ApJ:2003, Palmeri:AA:2003, Mendoza:AA:2004, Gorczyca:ApJ:2006, Hasoglu:ApJ:2006, Garcia:ApJS:2009, Palmeri:AA:2011, Palmeri:AA:2012, Kucas:ApJ:2015}. 
\citet{Gorczyca:ApJ:2003} found that their branching ratios are roughly in agreement with those of \citet{Kaastra:AAS:1993} for light ions, $Z \lesssim 20$, but there are significant discrepancies for heavier ones. An improved set of branching ratio data would be very useful, but for now \citet{Kaastra:AAS:1993} remains the only comprehensive dataset. In Section~\ref{sec:app}, we point out some specific instances where we find discrepancies that can be attributed to inaccurate branching ratios. 

\citet{Shevelko:JPhysB:2006} also presented experimental fits to double-ionization data for a few ions. These fits are more accurate than the general formula above. In this scheme the direct cross section is given by 
\begin{equation}
\sigma_{\mathrm{D}}=1-\mathrm{e}^{-3\left(u-1\right)} \frac{p_0}{E_{\mathrm{th}}^3}\left[\frac{u-1}{\left(u+0.5\right)^2}\right]\left[1+0.1\ln{(4u+1)}\right] 
\times10^{-13} \, \mathrm{cm^{2}},
\label{eq:shev6a}
\end{equation}
where $p_0$ is a fit parameter. The IA cross sections are given by 
\begin{equation}
\sigma_{\mathrm{IA}}=\frac{p_0}{E_{\mathrm{th}}^2}\frac{u-1}{u \left(u+5.0 \right)} \left[1+\frac{0.3}{p_{1}}\ln{(4u+1)}\right]
\times10^{-13} \, \mathrm{cm^{2}},
\label{eq:shev6b}
\end{equation}
where $p_0$ is a fit parameter and $p_1$ is the principal quantum number of the core electron that is directly ionized. Unlike for Equations~(\ref{eq:shev5a}) and (\ref{eq:shev5b}), the parameters for Equations~(\ref{eq:shev6a}) and (\ref{eq:shev6b}) do not depend on the isoelectronic sequence, but instead are set by fits to specific experiments. As a result, these formulae are not suitable for extension to unmeasured systems. 

In many cases, the IA process can be represented very well by the Lotz formula  \citep{Lotz:ZPhys:1969} for single ionization of a core electron multiplied by the branching ratio $f_{\mathrm{BR}}$ for autoionization of the resulting intermediate state, 
\begin{equation}
\sigma_{\mathrm{IA}}=4.5 f_{\mathrm{BR}} p_{0}\frac{\ln{(u)}}{E_{\mathrm{th}}^{2}u}
\times10^{-14} \, \mathrm{cm^{2}}.
\label{eq:lotz}
\end{equation}
Here $p_0$ is the initial number of electrons in the subshell where the ionization takes place and $E_{\mathrm{th}}$ is the threshold for single ionization of that subshell. The needed branching ratios for many ions of astrophysical interest have been given by \citet{Kaastra:AAS:1993}. As mentioned above, updated calculations for some systems have been reported. However, in most cases those calculations are less useful for estimating EIMI cross sections, because they report a total radiative yield, but do not give a breakdown of the branching ratios for each number of electrons ejected. 

Several other formulae have been proposed for representing EIMI cross sections. \citet{Fisher:JPhysB:1995} gave a set of two formulae that are relevant at low energies and high energies, respectively. As it is a somewhat more complicated scheme than any of the above formulae and the data can be adequately represented using a simpler approach, we have opted not to use the \citeauthor{Fisher:JPhysB:1995} formulae. \citet{Talukder:EurPhys:2009} also describe an approach for fitting double ionization cross sections for some light and low-charged ions, but their parameters are direct fits to experimental data and so cannot be directly extended along isoelectronic sequences.

\section{Fitting Formulae and Parameters by Isoelectronic Sequence}\label{sec:app}

We have fit the above cross section formulae to various reported EIMI measurements. Table~\ref{table:refs} gives the bibliographic reference for the experimental EIMI data, organized by isoelectronic sequence and order of EIMI (i.e., double, triple, quadruple, etc.). Although all of the above formulae have been developed or previously applied to some experimental data, our purpose in doing this new comparison was to determine: First, whether the formulae remained valid in light of data that have been published more recently. Second, under which conditions, such as isoelectronic sequence and nuclear charge $Z$, each semiempirical formula best represents the data. And third, to revise the parameters when necessary for our applications. In this section we report the results by isoelectronic sequence. 

In the following, the direct EIMI thresholds are from \citet{NIST:2016} and represent the sum of the single ionization thresholds from initial charge state $q_{i}$ to final charge state $q_{f}$. The IA thresholds are the single ionization thresholds for the core electrons and are taken from \citet{Kaastra:AAS:1993}, except for some light ions and low-charged ions of argon, for which more recent data are given in \citet{Shevelko:JPhysB:2005}. We also do not account for relativistic effects, which may be relevant as collision energies become significant compared to the electron rest mass. For single ionization, relativistic corrections become noticeable at about 20~keV \citep{Kim:PRA:2000}.

Table~\ref{table:cross} presents a sample of the online data table, which lists the parameters used for each cross section. In that table, each ion is denoted by the nuclear charge $Z$, initial charge state $q_{i}$, and final charge state $q_{f}$. The total cross section is the sum of all the individual cross sections having the same set of $Z$, $q_{i}$, and $q_{f}$. Also, although we previously reported EIMI cross section estimates for Fe \citep{Hahn:ApJ:2015}, here we have made some minor revisions to those cross sections based on the current analysis of a wider range of ions.

\subsection{He-like}\label{subsubsec:he}

\begin{figure}
\centering \includegraphics[width=0.9\textwidth]{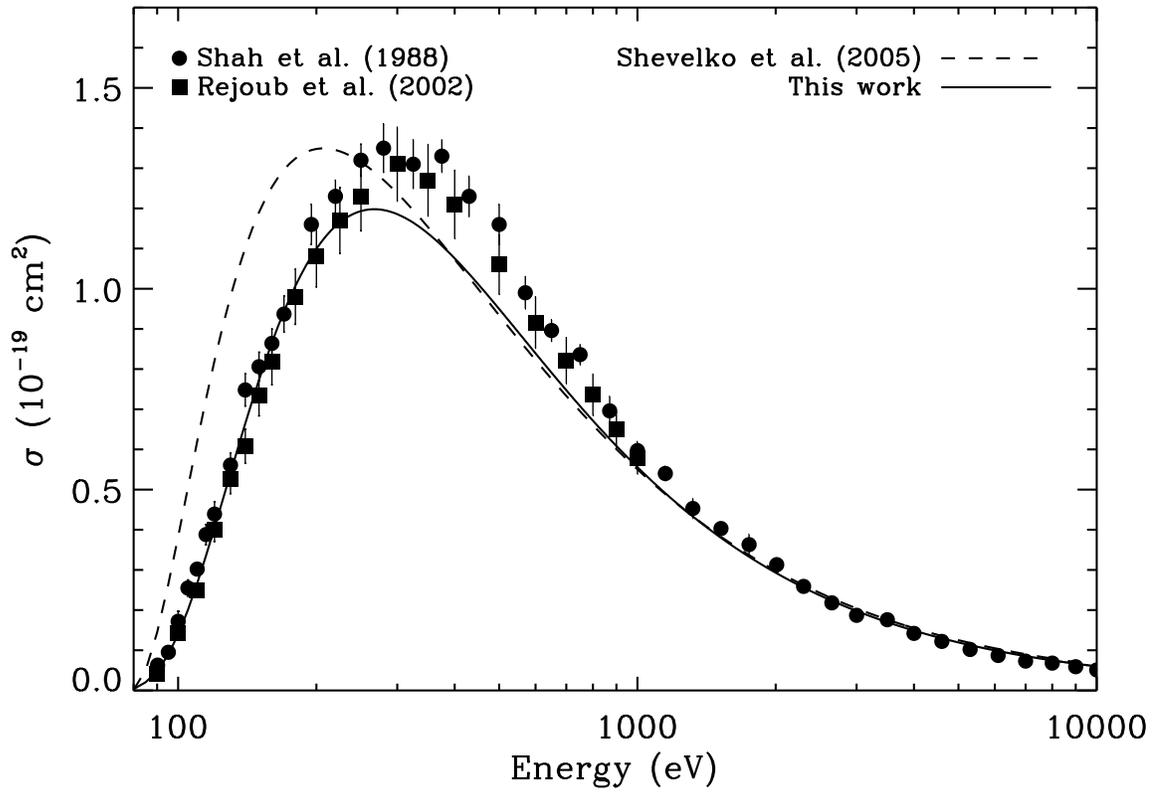}
\caption{\label{fig:he02} Double ionization of He$^{0+}$ forming He$^{2+}$. The data points show the experimental results of \citet[filled circles][]{Shah:JPhysB:1988} and \citet[filled squares][]{Rejoub:PRA:2002}. The dashed line illustrates the semiempirical formula following \citet{Shevelko:JPhysB:2005} and the solid line uses the cross-section scaling found here for He-like systems.
}
\end{figure}

He-like ions are difficult to fit using any of the semiempirical formulae described above. The dashed curve in Figure~\ref{fig:he02} illustrates the formula using Equation~(\ref{eq:shev5a}) from \citet{Shevelko:JPhysB:2005} compared to various double ionization measurements of neutral He \citep{Shah:JPhysB:1988, Rejoub:PRA:2002}. There is clearly an offset in the energy position of the peak cross section. Similar results are found for other He-like ions. 

\begin{figure}
\centering \includegraphics[width=0.9\textwidth]{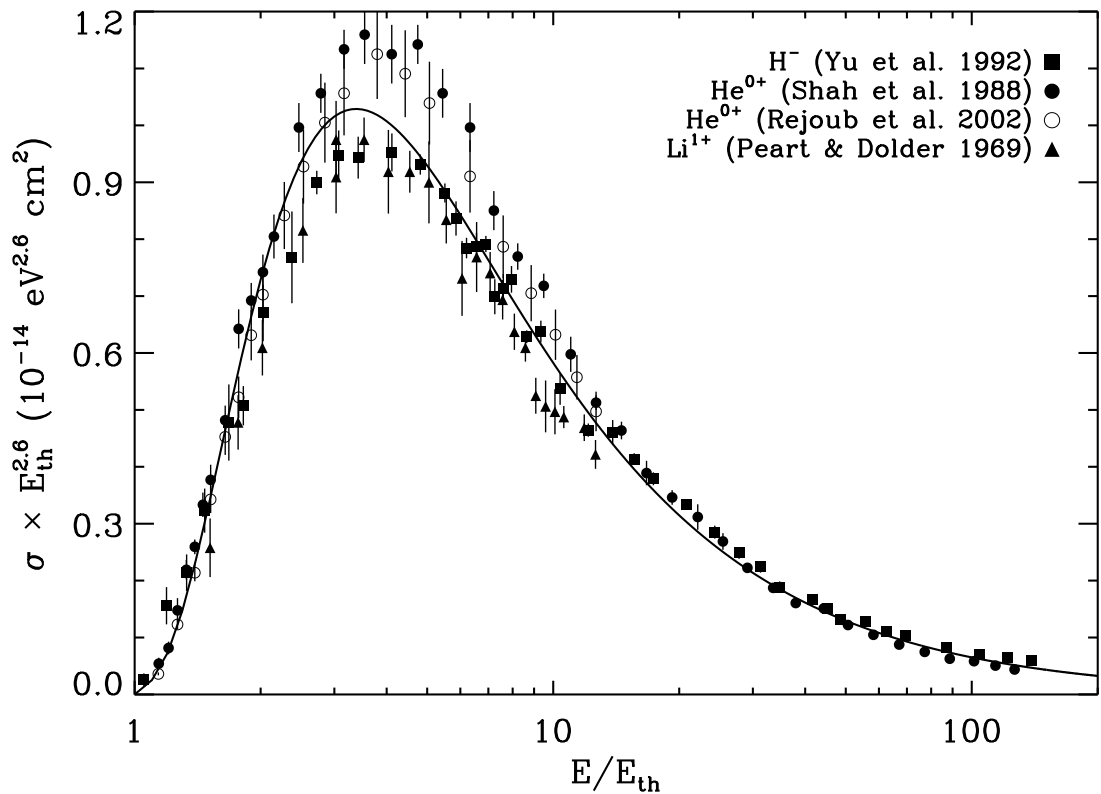}
\caption{\label{fig:hescale} Scaled double ionization cross sections for the He-like systems  H$^{-}$, He$^{0+}$, and Li$^{1+}$. The solid line shows our fit to the scaled data using Equation~(\ref{eq:alfred}). 
}
\end{figure}

In order to fit the He-like data, we compared double ionization cross section data for H$^{-}$ \citep{Yu:JPhysB:1992}, He$^{0+}$ \citep{Shah:JPhysB:1988,Rejoub:PRA:2002}, and Li$^{1+}$ \citep{Peart:JPhysB:1969}. Figure~\ref{fig:hescale} shows that if the cross sections are scaled by a factor of $E_{\mathrm{th}}^{-2.6}$ and plotted against the normalized energy $E/E_{\mathrm{th}}$, then the scaled cross sections agree very well \citep[see also][]{Muller:NIMB:2005}. \citet{Kim:PRA:1994} suggested that double ionization of He-like ions should be well fit by a function of the form:
\begin{equation}
\sigma = \frac{1}{E_{\mathrm{th}}^{p_{0}}} \frac{1}{u} 
\left[ p_{1}\left(1-\frac{1}{u}\right) + p_{2} \frac{\ln(u)}{u} + p_{3} \frac{\ln(u)}{u^2} + p_{4} \frac{\ln(u)}{u^3} \right] 
\times10^{-14} \, \mathrm{cm^{2}}.
\label{eq:alfred}
\end{equation}
Here, $p_{0}=2.6$ is the scaling factor described above. The parameters $p_{1}$--$p_{4}$ are determined from the fit to the scaled data and are given in Table~\ref{table:cross}. This function fits the scaled data very well as shown by Figure~\ref{fig:hescale}. The solid curve in Figure~\ref{fig:he02} shows our results applied to double ionization of He$^{0+}$. 

We extrapolate the He-like double ionization cross sections to unmeasured systems using Equation~(\ref{eq:alfred}). Note that all of the parameters $p_{0}$--$p_{4}$ are the same within the isoelectronic sequence, with the only change coming from the varying $E_{\mathrm{th}}$. Measurements of double ionization for heavier He-like ions would be useful in order to assess the accuracy of this extrapolation. 

For other isoelectronic sequences the peak of the direct double ionization cross section occurs roughly in the position predicted by the Equation~(\ref{eq:shev5a}).
That is, the formula as presented by \citet{Shevelko:JPhysB:2005} is adequate. The reason for the shift of the peak to higher energies for the He-like ions may be due to the strong interaction of the electrons with one another for these two-electron systems \citep{Defrance:JPhysB:2000, Defrance:NIMB:2003, Shevelko:JPhysB:2005}. Experimental measurements of the differential cross section for double ionization of helium also suggest that the dominant direct ionization mechanism is TS2, in which the incident electron collides sequentially with both bound electrons \citep{Lahmam:JPhysB:2010}.

\subsection{Li-like}\label{subsubsec:li}

Double ionization of Li-like ions is described well using Equation~(\ref{eq:shev5a}). For double ionization of Li ($Z=3$), we found that $p_{0}=5.8$ reproduces well the measured cross sections of \citet{Jalin:JChemPhys:1973} and \citet{Huang:PRA:2002}. However, the value $p_{0}=12$ given by \citet{Shevelko:JPhysB:2005} matches the measured cross sections for C$^{3+}$ and N$^{4+}$ of \citet{Westermann:PhysScr:1999} very well. Lacking any additional experimental data, for all cross sections with $Z \geq 4$ we set $p_{0}=12$. IA does not contribute because the resulting excited state lies below the ionization threshold. That is, a vacancy in the $K$-shell, is filled by the relaxation of the $2s$ electron. As there is no other electron in an outer shell that can absorb the energy, only radiative relaxation is possible. 

The only Li-like system for which triple ionization data are available is Li, for which \citet{Huang:PRL:2003} measured the ratio of triple to double ionization at 1000~eV to be about $1 \times 10^{-3}$. To match the cross section implied by this ratio, we have rescaled Equation~(\ref{eq:shevtar}) by $f=0.003$, while otherwise using the triple ionization parameters given by \citet{Shevelko:JPhysB:1995}. This is a very large discrepancy for the unaltered prediction formula and is possibly due to the formula having been developed on the basis of data for ions with many more electrons. We ignore triple ionization of Li-like ions for higher $Z$ ions, because the cross section for Li is already very small and is expected to decrease strongly with increasing $Z$. Data are lacking for other ions. 

\subsection{Be-like}\label{subsubsec:be}

\begin{figure}
	\centering \includegraphics[width=0.9\textwidth]{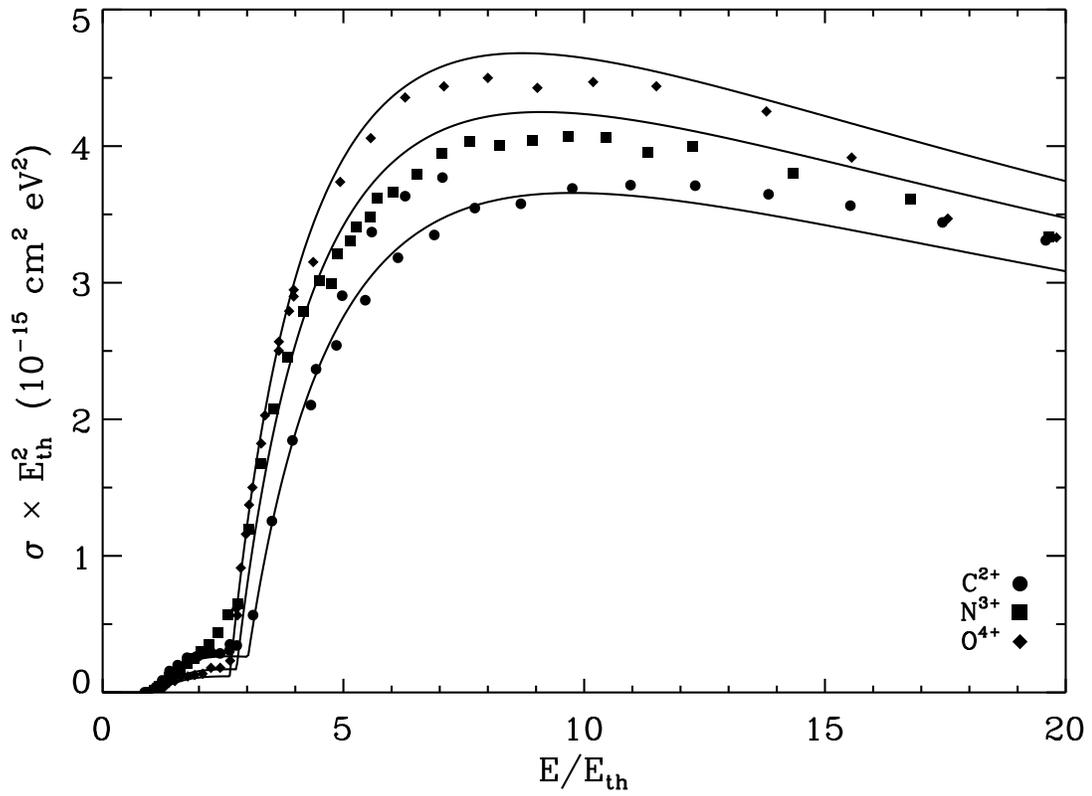}
	\caption{\label{fig:belike} Scaled double ionization cross sections for several Be-like ions: C$^{2+}$, N$^{3+}$, and O$^{4+}$. The symbols show the measurements of \citet{Westermann:PhysScr:1999} and the solid curves are the cross sections reported here, which in this case use the semiempirical scheme of \citet{Shevelko:JPhysB:2005}. 
	}
\end{figure}

For this sequence, double ionization measurements exist for B$^{1+}$, C$^{2+}$, N$^{3+}$, O$^{4+}$, and Ne$^{6+}$. All of these appear to be well-matched by summing Equations~(\ref{eq:shev5a}) and (\ref{eq:shev5b}) with the parameters for the direct double ionization given by \citet{Shevelko:JPhysB:2005}. Examples of these fits for C$^{2+}$, N$^{3+}$, and O$^{4+}$ are shown in Figure~\ref{fig:belike} and compared to the data of \citet{Westermann:PhysScr:1999}. For the figure, the cross section data have been scaled by a factor of $E_{\mathrm{th}}^2$. As the charge state increases, IA becomes increasingly important relative to direct ionization. \citet{Kaastra:AAS:1993} give the branching ratio for IA of Be-like ions as unity, however the calculations of \citet{Shevelko:JPhysB:2005} suggest that for high $Z$ ions, the branching ratio is closer to 90\%. We are not aware of any experimental data for systems above $Z=10$ that we could compare to, but this would be a relatively modest error. On the basis of these data, we infer the cross sections for all other Be-like ions using Equations~(\ref{eq:shev5a}) and (\ref{eq:shev5b}) with branching ratios from \citet{Kaastra:AAS:1993}.

Theoretical calculations have been performed for double ionization of both Be$^{0+}$ and B$^{1+}$ using time-dependent close-coupling $R$-matrix and distorted-wave methods \citep{Pindzola:JPhysB:2010, Pindzola:JPhysB:2011}. Their calculations for B$^{1+}$ are in good agreement with the experimental data. However, for Be$^{0+}$ there are no experimental data with which to compare.  The semiempirical prediction for direct ionization of Be$^{0+}$ is about 50\% larger than predicted by the theory, but the larger IA contributions appear to be roughly in agreement. Given that the theory is accurate for B$^{1+}$ and that the semiempirical formula seems to overestimate the direct double ionization cross section for very low charge states in nearby isoelectronic sequences (e.g., Li-like  and B-like), it is possible that the Be$^{0+}$ theory is correct and that the semiempirical formula is an overestimate here as well. Experimental measurements would be beneficial to resolve this discrepancy, but for now we use the semiempirical prediction.

Because the state resulting from a $K$-shell vacancy lies below the double ionization threshold for the resulting system, net triple and quadruple ionization of Be-like ions can only proceed via direct ionization processes. As there are no measurements and we expect these cross sections to be very small, as they are for triple ionization of Li (Section~\ref{subsubsec:li}), we ignore them.

\subsection{B-like}\label{subsubsec:b}

Double ionization of B-like ions is generally well-described using Equations~(\ref{eq:shev5a}) and (\ref{eq:shev5b}). For direct ionization using Equation~(\ref{eq:shev5a}), \citet{Shevelko:JPhysB:2005} recommend $p_{0}=10$, which matches the experiments for N$^{2+}$, O$^{3+}$, and Ne$^{5+}$, but not the data for C$^{1+}$. For C$^{1+}$, we set $p_{0}=8.24$ to match the cross section data of \citet{Westermann:PhysScr:1999}. For DI of all the other ions we set $p_{0}=10$. Based on our results for other near-neutral species, this may overestimate the DI cross section for B$^{0+}$. 

IA contributes to the double ionization of B-like ions and may also contribute to triple ionization. The predicted IA branching ratios for B-like ions can be benchmarked by comparison to the experiments. \citet{Kaastra:AAS:1993} write that following ionization of a 1$s$ electron the B-like ions either radiatively stabilize for a net single ionization or eject one electron leading to a net double ionization. They predict that the branching ratio for triple ionization is zero. However, measurements of triple ionization for C$^{1+}$, N$^{2+}$, and O$^{3+}$ by \citet{Westermann:PhysScr:1999} suggest that IA does contribute to triple ionization. These cross sections can be well-described by the Lotz cross section (Equation~\ref{eq:lotz}) for single ionization of the $1s$ electron multiplied by a factor of 0.068, 0.04, or 0.024, respectively. This suggests that the branching ratio for triple ionization has a magnitude of a few percent. 

Another indication that IA leads to triple ionization comes from the double ionization measurements of Ne$^{5+}$ by \citet{Duponchelle:JPhysB:1997}. These double ionization data could be better fit by Equation~(\ref{eq:shev5b}) if the IA branching ratio were reduced by 0.1, from the \citeauthor{Kaastra:AAS:1993} prediction of 0.98 to about 0.88. The reduction suggests that Ne$^{5+}$ could have an IA branching ratio for triple ionization of up to $\sim 0.1$. 

Triple ionization of B-like C, N, and O are best described as being due solely to IA, using the Lotz formula (Equation~\ref{eq:lotz}) multiplied by the empirical branching ratios given above. We use the same method to estimate the cross section for triple ionization of Ne$^{5+}$, although, as discussed above, the estimate for that ion is not based directly on triple ionization measurements but rather inferred from the double ionization cross section. For other B-like ions we assume that IA will continue to dominate the cross section and adopt an estimated branching ratio of $f_{\mathrm{BR}}=0.06$, which is the average of branching ratios we estimated above based on the C, N, O, and Ne data. For consistency, we also reduce the corresponding branching ratios for double ionization by $0.06$ to account for the fraction that we now ascribe to triple ionization.

We ignore quadruple and higher order EIMI processes, as we are unaware of any measurements beyond triple ionization, and such cross sections are likely to be very small as they would require breaking open the $K$-shell.

\subsection{C-like}\label{subsubsec:c}

Double and triple ionization of C-like ions are estimated in a similar way as for B-like ions. For double ionization, \citet{Shevelko:JPhysB:2005} sets $p_{0}=23$ in Equation~(\ref{eq:shev5a}). This is accurate compared to measurements of O$^{2+}$ and Ne$^{4+}$, but we reduce to $p_{0}=19.2$ in order to match the double ionization measurements of N$^{1+}$ by \citet{Lecointre:JPhysB:2013}. Nitrogen measurements were also reported by \citet{Zambra:JPhysB:1994}, but their cross section is about 30\% smaller than that of \citet{Lecointre:JPhysB:2013}. We choose not to rely on the \citet{Zambra:JPhysB:1994} results, because they appear to be systematically smaller than the results of other groups.  Triple ionization measurements for N and O suggest that the IA branching ratio for triple ionization is about 0.07 and 0.075, respectively and so we reduce \citet{Kaastra:AAS:1993} IA branching ratio for double ionization by these amounts in order to account for the fraction of IA that leads to triple rather than double ionization. 

The triple ionizaton measurements for N and O are best modeled as dominated by IA using the Lotz formula (Equation~\ref{eq:lotz}). Based on these results, we estimate the cross section for other C-like ions in the same way, adopting a branching ratio of $f_{\mathrm{BR}}=0.07$. Consistent with this representation for triple ionization, we reduce the branching ratios for double ionization by 0.07 from the \citet{Kaastra:AAS:1993} values.

Quadruple ionization of N$^{1+}$ was measured by \citet{Lecointre:JPhysB:2013}. None of the formulae seem to match the data very well, but the data are also not very detailed with large statistical error bars. Assuming that the cross section is due to $1s$ IA following Equation~(\ref{eq:lotz}), a branching ratio for quadruple ionization of $1.4\times10^{-3}$ would approximate the magnitude of the cross section, but the energy threshold is too low. An alternative is the indirect MIMA process in which multiple direct ionization of a $1s$ and a $2s$ electron leads to a system that subsequently relaxes through double autoionization giving a net quadruple ionization. Using the Los Alamos Atomic Physics Code \citep{Magee:ASP:1995}, we estimate that the threshold for this process is about 496~eV, which would match the apparent threshold in the data quite well. Based on this, we model the cross section as direct double ionization of a $1s$ and $2s$ electron pair using Equation~(\ref{eq:shevtar}) for double ionization with a threshold of $\approx496$~eV and scaled by a factor of $f=0.0225$ to match the data. This $f$ could be interpreted as the branching ratio for quadruple ionization after forming the two ``holes'' in the $1s$ and $2s$ subshells. However, given that the Equation~(\ref{eq:shevtar}) can sometimes be very inaccurate for predicting double ionization, the inferred branching ratio should be regarded as an order of magnitude estimate, at best.

For estimating quadruple ionization of other C-like ions, we scale Equation~(\ref{eq:shevtar}) by a factor of $f=0.1$. This scaling is based on considering what is needed to match the magnitude of the quadruple ionization cross sections for C-, N-, and Ne-like ions. For example, scaling Equation~(\ref{eq:shevtar}) by a factor of $f\approx0.04$ would roughly match the peak of the quadruple ionization cross section for N$^{1+}$, while factors of $f\approx 0.06$ and $f\approx0.12$ would suffice for quadruple ionization of N-like O$^{1+}$ and Ne-like Ne$^{0+}$, respectively. We apply this approach for estimating the quadruple ionization cross section for all ions that lack experimental data in the C-like through Ne-like isoelectronic sequences. This approximation probably underestimates the effective energy threshold for quadruple ionization, because the thresholds for the important indirect processes lie at higher energies than the DI threshold used in Equation~(\ref{eq:shevtar}). EIMI beyond quadruple ionization is ignored for C-like ions.

\subsection{N-like}\label{subsubsec:n}

Double ionization of N-like ions O$^{1+}$, Ne$^{3+}$, and Ar$^{11+}$ is well-described by Equations~(\ref{eq:shev5a}) and (\ref{eq:shev5b}) using the parameters given by \citet{Shevelko:JPhysB:2005} and the branching ratios for IA from \citet{Kaastra:AAS:1993}. We apply the same scheme to double ionization for all other N-like ions. 
 
Triple and quadruple ionization measurements exist for O$^{1+}$ \citep{Westermann:PhysScr:1999, Lecointre:JPhysB:2013}. For triple ionization the data can be best modeled by scaling Equation~(\ref{eq:shevtar}) by a factor of $f=0.056$ with the other parameters given by \citet{Shevelko:JPhysB:1995} and adding to that an IA contribution modeled using Equation~(\ref{eq:lotz}) for the K-shell ionization with a branching ratio of $f_{\mathrm{BR}}=0.108$. This inferred triple ionization branching ratio is roughly similar to that found in photoionization measurements of N$^{0+}$ by \citet{Stolte:ApJ:2016}. The direct and indirect contributions to triple ionization are of similar magnitude. For triple ionization of other N-like ions, we scale Equation~(\ref{eq:shevtar}) by a factor of $f=0.15$, which has the observed threshold for direct triple ionization and roughly matches the magnitude of the total cross section for N-like ions, as well as for nearby isoelectronic sequences. There is not sufficient data to reliably separate the direct and IA contributions.

\begin{figure}
	\centering \includegraphics[width=0.9\textwidth]{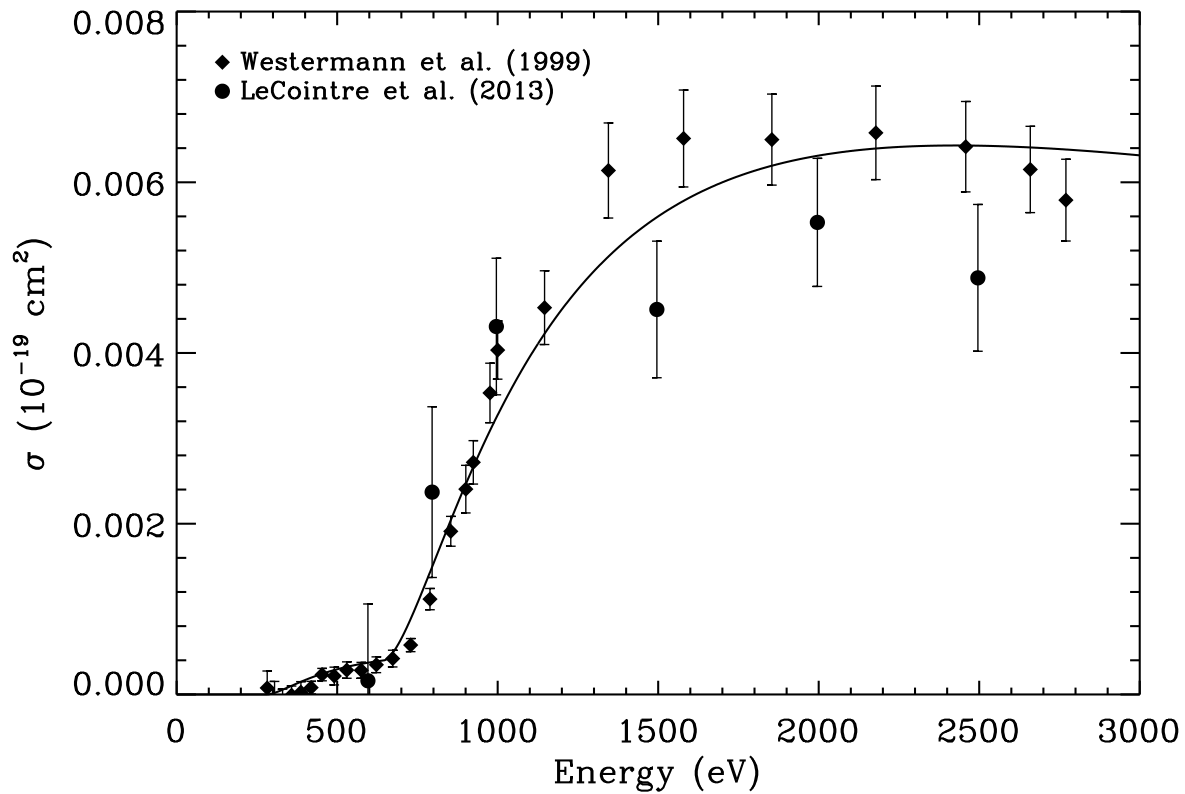}
	\caption{\label{fig:o15} Quadruple ionization of O$^{1+}$ forming O$^{5+}$. The filled diamonds show the measurements of \citet{Westermann:PhysScr:1999} and the filled circles indicate those of \citet{Lecointre:JPhysB:2013}. The solid curve is our model of the data, as described in Section~\ref{subsubsec:n}. 
	}
\end{figure}

Figure~\ref{fig:o15} shows the quadruple ionization cross section for O$^{1+}$. Similar to N$^{1+}$, this can be modeled as the sum of direct quadruple ionization, plus an indirect cross section due to MIMA, where direct ionization of both a $1s$ and a $2s$ electron results in a system that relaxes through further double autoionization. The data show that the cross section begins to rise at the direct quadruple ionization threshold of about $281$~eV and that this appears to be the only contribution to the cross section until $\sim650$~eV. This direct cross section is well-matched by Equation~(\ref{eq:shevtar}) scaled by $f=0.005$. In principle, we might expect IA from the $K$-shell to lead to quadruple ionization, but no increase in the cross section is seen until about 100~eV above the $K$-shell ionization threshold of $\approx568$~eV. Using the Los Alamos Atomic Physics Code \citep{Magee:ASP:1995}, we estimate that the threshold for double direct ionization of a $1s$ and $2s$ electron pair is $\approx646$~eV, which matches the threshold in the data. We estimate the MIMA cross section using Equation~(\ref{eq:shevtar}) for double ionization of the $1s$ and $2s$ electron pair scaled by a factor of $f=0.078$. This scaling factor can be thought of as the branching ratio for the system to relax by double autoionization. However, it is better to consider it a fitting parameter only, because for it to be a quantitative estimate of the branching ratio requires that Equation~(\ref{eq:shevtar}) accurately estimate the inner-shell direct double ionization cross section, which may not be the case.

Although MIMA appears to be the dominate the O$^{1+}$ quadruple ionization cross section, there are not sufficient data to extrapolate its contribution to other N-like systems. Instead, we estimate quadruple ionization for other N-like ions by using Equation~(\ref{eq:shevtar}) scaled down by a factor of $f=0.1$, as described above in Section~(\ref{subsubsec:c}). 

\subsection{O-like}\label{subsubsec:o}

\begin{figure}
	\centering \includegraphics[width=0.9\textwidth]{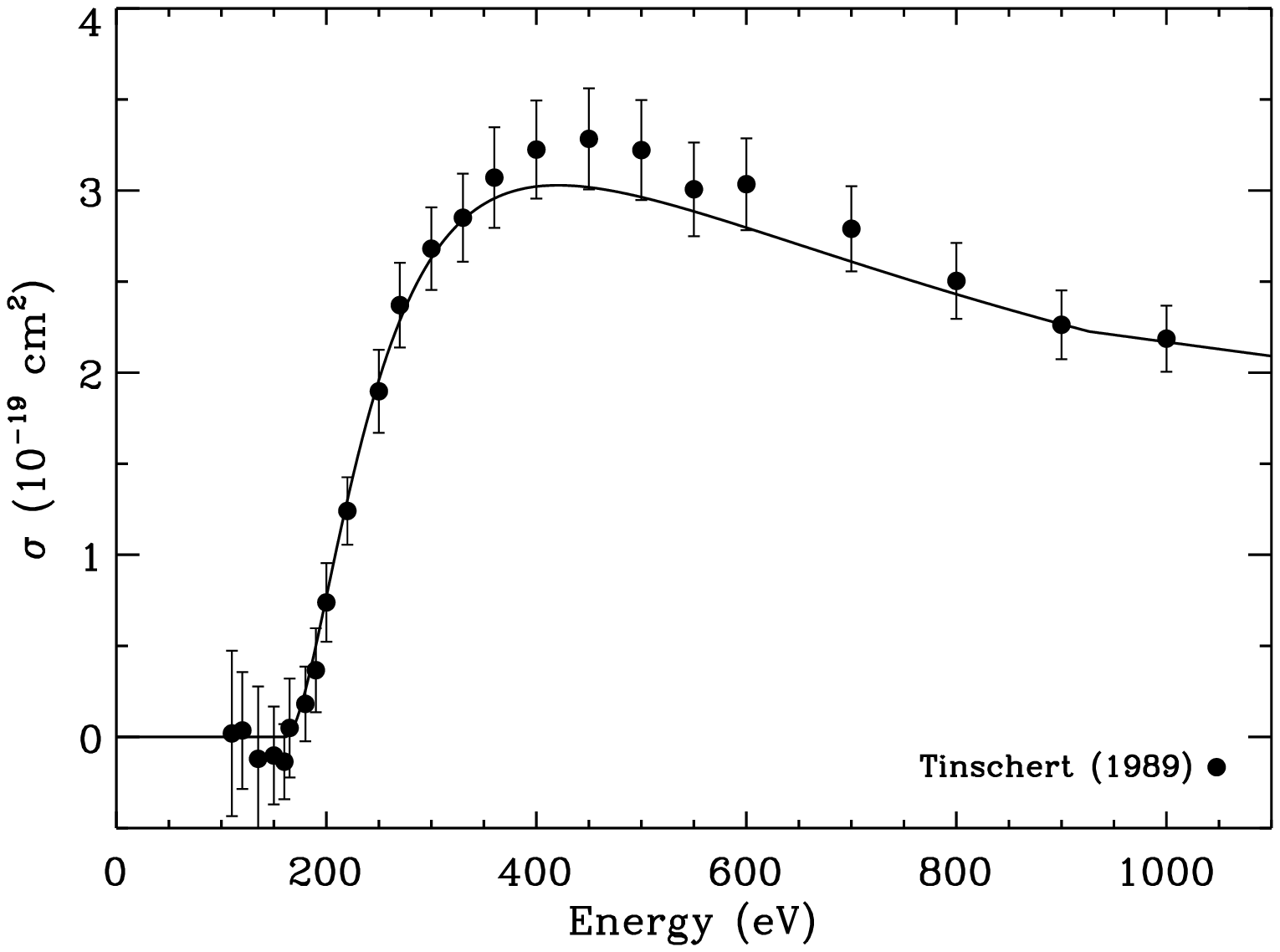}
	\caption{\label{fig:ne24} Double ionization of Ne$^{2+}$ forming Ne$^{4+}$. The symbols show the measurements of \citet{Tinschert:1989} and the solid curve is our fit to the data as described in Section~\ref{subsubsec:o}.
	}
\end{figure}

\begin{figure}
	\centering \includegraphics[width=0.9\textwidth]{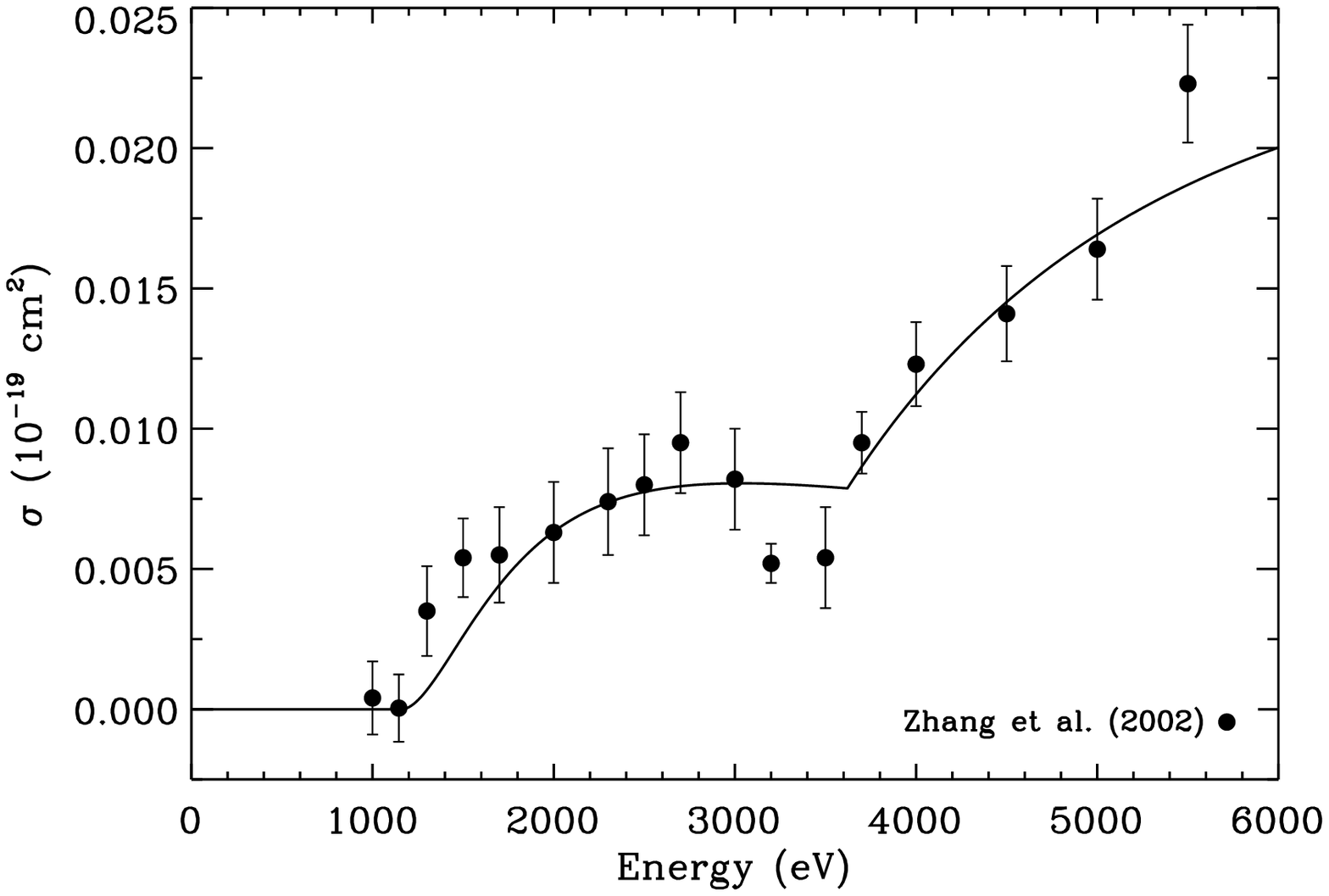}
	\caption{\label{fig:ar1012} Double ionization of Ar$^{10+}$ forming Ar$^{12+}$. The symbols show the measurements of \citet{Zhang:JPhysB:2002} and the solid curve is our fit to the data as described in Section~\ref{subsubsec:o}.
}
\end{figure}

Equations~(\ref{eq:shev5a}) and (\ref{eq:shev5b}), with parameters from \citet{Shevelko:JPhysB:2005} and IA branching ratios from \citet{Kaastra:AAS:1993} match the experimental measurements of double ionization of O-like ions F$^{1+}$, Ne$^{2+}$ (Figure~\ref{fig:ne24}), and Ar$^{10+}$ (Figure~\ref{fig:ar1012}) well. However, the same procedure overestimates the cross section for O$^{0+}$ by about a factor of two. To correct for that, we set $p_{0}=39$ in Equation~(\ref{eq:shev5a}) for double ionization of neutral oxygen. For all other ions we make no modifications. 

We are not aware of any triple or higher order EIMI measurements for these ions. On the basis of results for nearby isoelectronic sequences, we model triple and quadruple ionization using Equation~(\ref{eq:shevtar}) scaled by a factor of $f=0.15$ and $f=0.1$, respectively. 

\subsection{F-like}\label{subsubsec:f}

Double ionization measurements exist for Ne$^{1+}$ \citep{Tinschert:1989, Zambra:JPhysB:1994}, Al$^{4+}$ \citep{Steidl:1999}, and Ar$^{9+}$ \citep{Zhang:JPhysB:2002}. We find that Equations~(\ref{eq:shev5a}) and (\ref{eq:shev5b}) can match the measured cross sections well. However, for neon we must set $p_{0}=76.7$ in Equation~(\ref{eq:shev5a}) to match the data, which is significantly smaller than the recommended value of $p_{0}=133$ \citep{Shevelko:JPhysB:2005}. The unmodified parameters do describe the aluminum and argon data well. 
	
\begin{figure}
	\centering \includegraphics[width=0.9\textwidth]{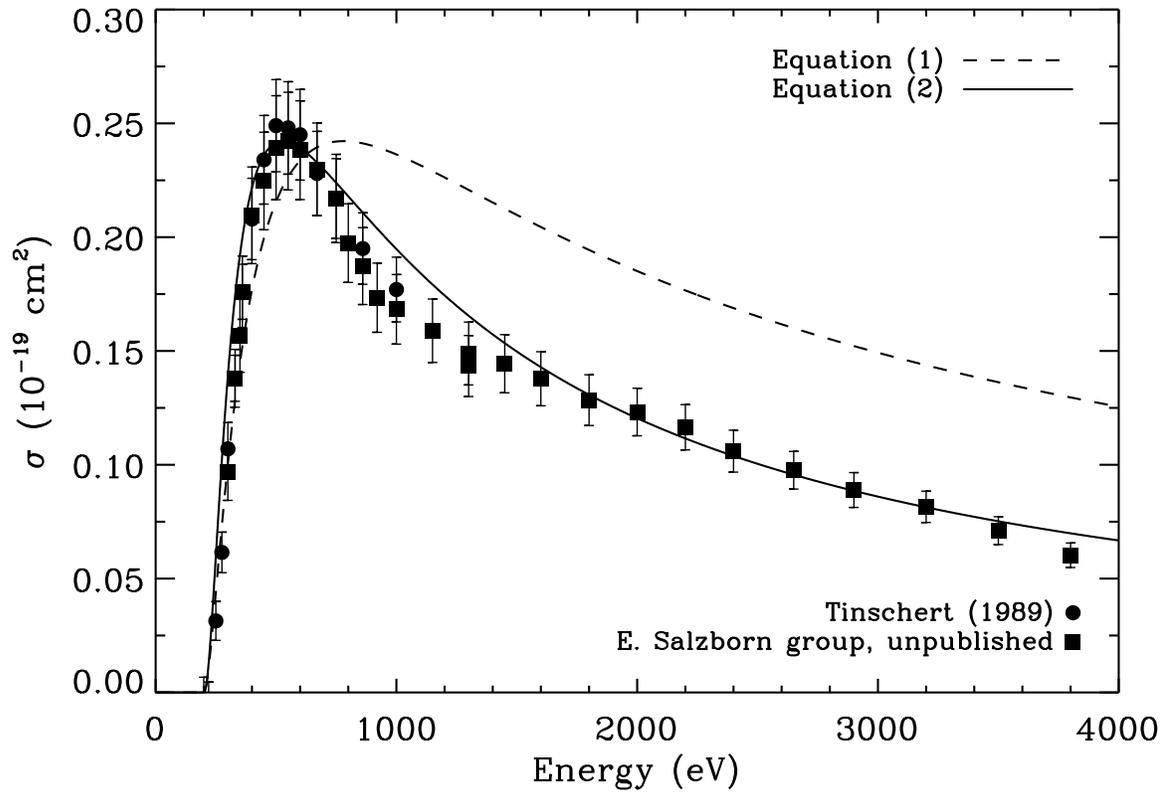}
	\caption{\label{fig:ne14} Triple ionization of Ne$^{1+}$ forming Ne$^{4+}$. The filled circles show the data of \citet{Tinschert:1989} and the filled squares indicate other unpublished data from the group of E. Salzborn (private communication). The dashed curve uses Equation~(\ref{eq:shevtar}), but the peak is clearly broader than the measurement. Instead we have fit the data using Equation~(\ref{eq:shev5a}). Although developed for double ionization cross sections, it fits these triple ionization data well. 
	}
\end{figure}

Figure~\ref{fig:ne14} shows the triple ionization cross section for Ne$^{1+}$ forming Ne$^{4+}$, which was measured by Tinschert, M\"uller, Hofmann and Salzborn \citep[][ unpublished]{Tinschert:1989}. 
There is also a separate unpublished measurement from the same group. Generally we fit triple ionization cross sections using Equation~(\ref{eq:shevtar}) for DI and model the IA using the Lotz cross section. However, for Ne$^{1+}$ the peak of the cross section is narrower than implied by Equation~(\ref{eq:shevtar}), so we use Equation~(\ref{eq:shev5a}) instead and set $p_{0}=12$ to match the experiment. An inflection is visible in the cross section at about 1000~eV that suggests an indirect ionization process, but the indirect contribution is relatively small and so we neglect it for this ion. 

For triple ionization of F-like ions other than Ne$^{1+}$ and quadruple ionization of all F-like ions, we use the same interpolation procedure as for other nearby isoelectronic sequences:  Equation~(\ref{eq:shevtar}) multiplied by factors of $f=0.15$ for triple ionization and $f=0.1$ for quadruple ionization. If we were to use this scheme to estimate the Ne$^{1+}$ triple ionization data, we would find that it is somewhat inaccurate as it overestimates the cross section at higher energy. Unfortunately, there are not enough data to test an alternative model.

\subsection{Ne-like}\label{subsubsec:ne}

There are several measurements for EIMI of neutral Ne$^{0+}$ forming Ne$^{2+}$ - Ne$^{5+}$ (see Table~\ref{table:refs}). Equation~(\ref{eq:shev5a}) can accurately match the double ionization measurements for Ne$^{0+}$ if $p_{0}=48.4$, which is significantly smaller than the recommended value of $p_{0}=183$ \citep{Shevelko:JPhysB:2005}.

IA is also expected to contribute to double ionization and can be included via Equation~(\ref{eq:shev5b}). \citet{Kaastra:AAS:1993} give a branching ratio of $f_{\mathrm{BR}}\approx0.98$ for net double ionization of Ne$^{0+}$, and they predict no contribution to triple or higher EIMI due to IA. \citet{Muller:ApJ:2017} recently studied photoionization of Ne atoms and Ne$^{1+}$ ions and measured the $K$-shell branching ratio to be $f_{\mathrm{BR}}=0.93$ for double ionization of Ne$^{0+}$ and we use this value to calculate the cross section.  $f_{\mathrm{BR}}\approx0.76$.

Direct triple ionization of neon is also well fit using Equation~(\ref{eq:shev5a}) with $p_{0}=16.6$, where again we have used this semiempirical formula that is normally intended for double ionization because it matches better the sharper measured peak than does Equation~(\ref{eq:shevtar}). The IA contribution is represented using Equation~(\ref{eq:shev5b}). \citet{Muller:ApJ:2017} found a branching ratio of $f_{\mathrm{BR}}=0.054$, which fits these data reasonably well. 
As usual, we are assuming that the only relevant indirect process is the single ionization of a $1s$ electron with subsequent Auger decays. 

\begin{figure}
	\centering \includegraphics[width=0.9\textwidth]{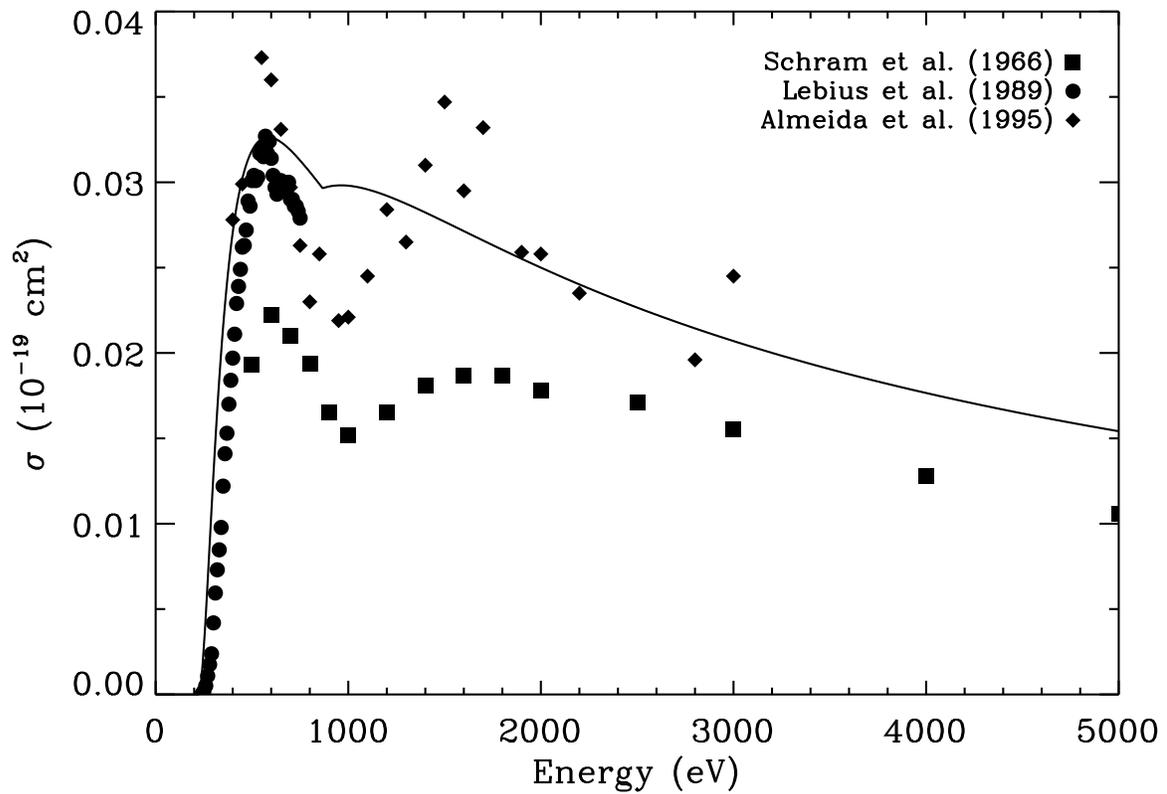}
	\caption{\label{fig:ne04} Quadruple ionization of Ne$^{0+}$ forming Ne$^{4+}$. The filled squares show the measurements of \citet{Schram:Physica:1966a}, the filled circles those of \citet{Lebius:JPhysB:1989}, and the filled diamonds those of \citet{Almeida:JPhysB:1995}. The solid curve is our model of the data as described in Section~\ref{subsubsec:ne}.
	}
\end{figure}

\begin{figure}
	\centering \includegraphics[width=0.9\textwidth]{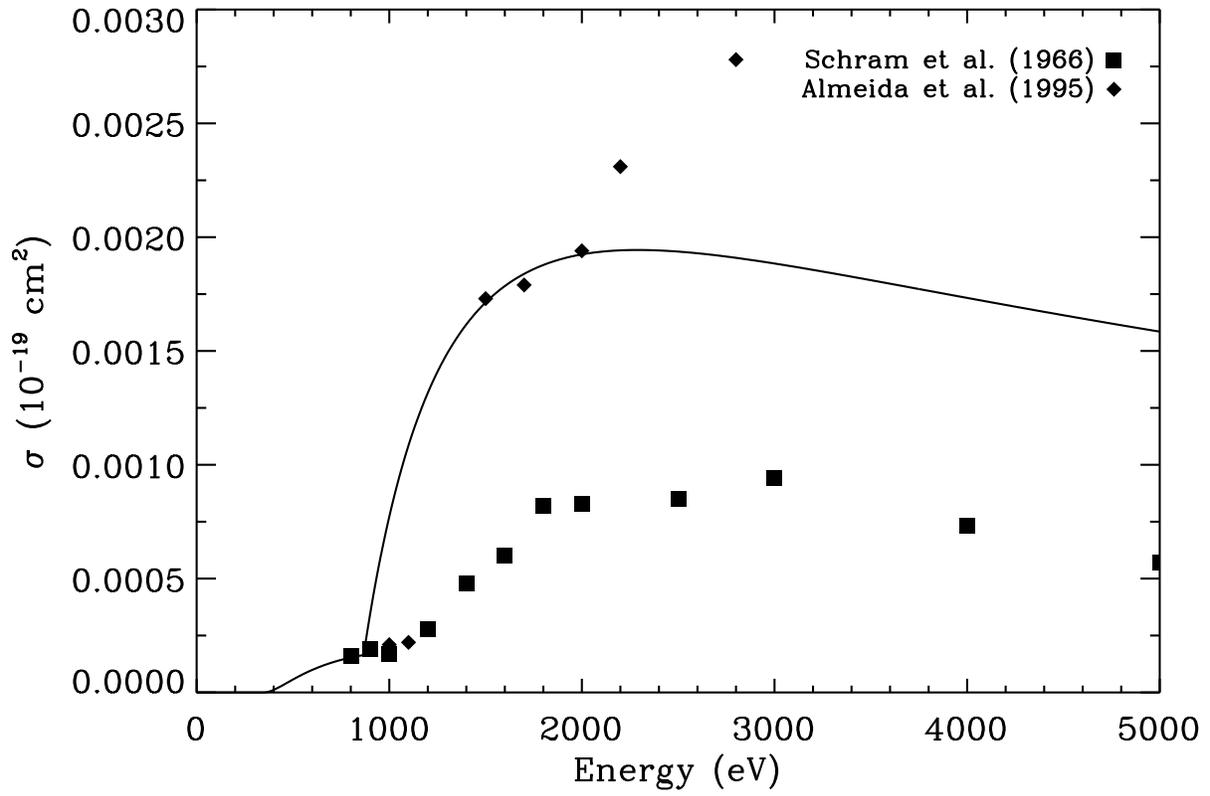}
	\caption{\label{fig:ne05} Quintuple ionization of Ne$^{0+}$ forming Ne$^{5+}$. The filled squares show the measurements of \citet{Schram:Physica:1966a} and the filled diamonds those of \citet{Almeida:JPhysB:1995}. The solid curve is our model of the data as described in Section~\ref{subsubsec:ne}.
	}
\end{figure}

The contribution of indirect processes is clear in the Ne$^{0+}$ quadruple and quintuple ionization cross sections, shown in Figures~\ref{fig:ne04} and \ref{fig:ne05}, respectively. There, the measurements imply branching ratios of $\approx 0.02$ for quadruple ionization and $\approx 0.004$ for quintuple ionization. This quadruple ionization branching ratio is an order of magnitude larger than that measured due to photoionization by \citet{Muller:ApJ:2017}, who found $f_{\mathrm{BR}}=0.002$. The discrepancy is possibly due to the contribution to the EIMI cross section of other indirect ionization processes, such as MIMA. We should also note that for quintuple ionization, there is a very large discrepancy of more than a factor of two between the measurements of \citet{Schram:Physica:1966a} and \citet{Almeida:JPhysB:1995}. The \citet{Schram:Physica:1966a} results appear to be systematically low for all of the neon EIMI data (compare Figures~\ref{fig:ne04} and \ref{fig:ne05}), so we base the branching ratio above on the \citet{Almeida:JPhysB:1995} measurements. We describe the direct quadruple ionization cross section using Equation~(\ref{eq:shev5a}) with $p_{0}\approx2.2$ and the direct quintuple ionization is modeled with Equation~(\ref{eq:shevtar}) scaled by a factor of $f=0.005$ to match the cross section below the IA threshold. 

Returning to double ionization of other Ne-like ions, measurements exist for Na$^{1+}$, Al$^{3+}$, and Ar$^{10+}$. We fit all of these data using Equation~(\ref{eq:shev5a}) for the direct ionization and Equation~(\ref{eq:shev5b}) for the indirect contribution. For the direct contribution, we find that $p_{0}$ often differs from the recommended value from \citet{Shevelko:JPhysB:2005} of $p_{0}=183$, with $p_{0}=91.5$ for Na$^{1+}$, 183 for Al$^{3+}$, and 201.6 for Ar$^{10+}$. The discrepancy is largest for low charge states and matches the nominal value for higher charge states. This is consistent with what we have seen for other isoelectronic sequences. Based on these measurements, we estimate the double ionization of the unmeasured Ne-like ions using the same equations with the unmodified $p_{0}$ value from \citet{Shevelko:JPhysB:2005} for Equation~(\ref{eq:shev5a}) and the IA branching ratios from \citet{Kaastra:AAS:1993} in Equation~(\ref{eq:shev5b}). 

We estimate the cross section for triple and quadruple ionization of $Z \geq 11$ Ne-like ions using Equation~(\ref{eq:shevtar}). Scaling factors of $f=0.4$ and $f=0.12$ provide a reasonable match to the magnitude of the Ne$^{0+}$ triple and quadruple ionization cross sections, respectively. The triple ionization of Al$^{3+}$ is well described by Equation~(\ref{eq:shevtar}) with $f=0.13$. Based on these results and on the data for nearby isoelectronic sequences, we set $f=0.15$ and $f=0.1$ for triple and quadruple ionization of Ne-like ions. We ignore quintuple ionization for other Ne-like ions, because there is not enough data to justify an extrapolation and because the cross section is very small.

\subsection{Na-like}\label{subsubsec:na}

\begin{figure}
	\centering \includegraphics[width=0.9\textwidth]{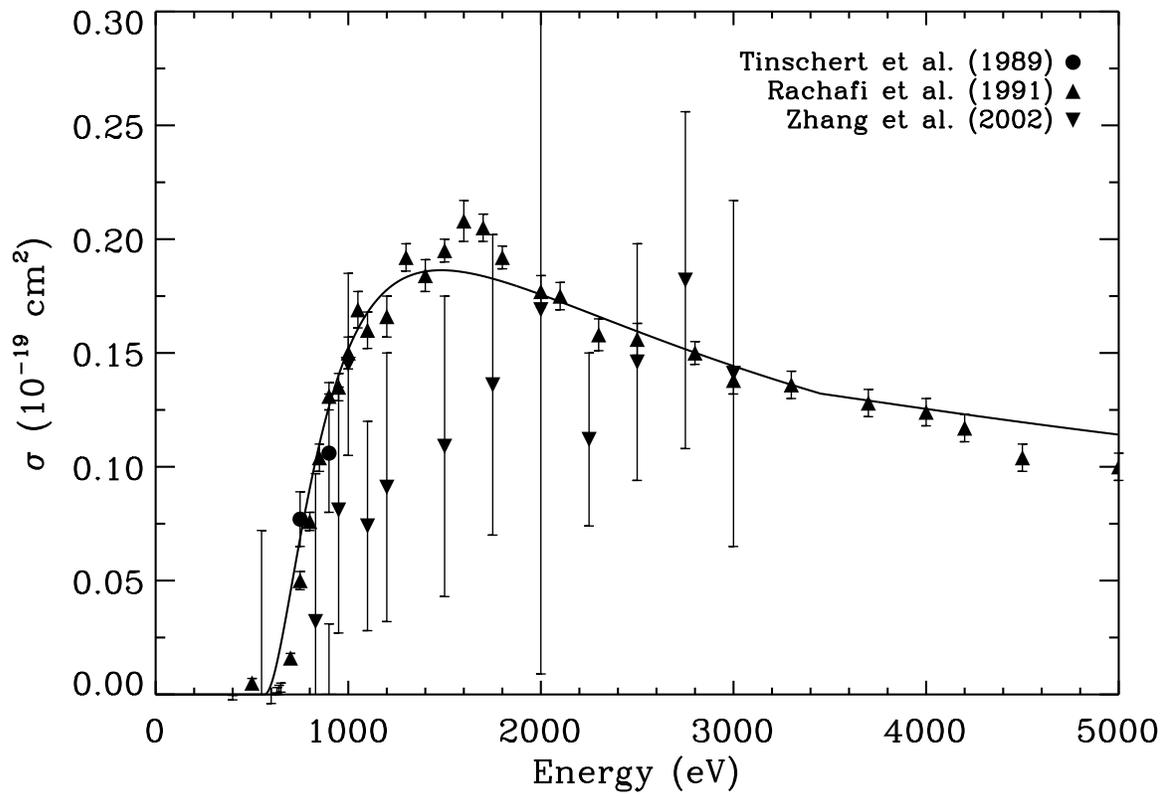}
	\caption{\label{fig:ar79} Double ionization of Ar$^{7+}$ forming Ar$^{9+}$. The filled circles show the data of \citet{Tinschert:JPhysB:1989}, the up-pointing triangles the data of \citet{Rachafi:JPhysB:1991} and the down-pointing triangles the data of \citet{Zhang:JPhysB:2002}. The solid curve illustrates our fit of the data using Equations~(\ref{eq:shev5a}) and (\ref{eq:shev5b}). 
	}
\end{figure}

Double ionization measurements exist for Al$^{2+}$ \citep{Steidl:1999} and for Ar$^{7+}$ \citep{Tinschert:JPhysB:1989, Rachafi:JPhysB:1991, Zhang:JPhysB:2002}. The Ar$^{7+}$ measurements are illustrated in Figure~\ref{fig:ar79}. For Ar$^{7+}$, the direct ionization contribution can be matched well using Equation~(\ref{eq:shev5a}) setting $p_{0}\approx204.7$. This is a factor of 10 smaller than the value of $p_{0}=2000$ recommended by \citet{Shevelko:JPhysB:2005}. However, they also based their parameters on the argon data, so we believe this is simply a typo in that work. The same formula with $p_{0}=200$ also matches the Al$^{2+}$ double ionization data well near the ionization threshold, although it underestimates the cross section at the peak and higher energies by about 30\%. Based on these results, we estimate the direct double ionization of other Na-like ions by setting $p_{0}=200$. For the indirect ionization of Na-like ions, including Ar$^{7+}$, we use Equation~(\ref{eq:shev5b}) with the branching ratio from \citet{Kaastra:AAS:1993}. There is also a double ionization measurement of Na$^{0+}$ by \citet{Tate:PR:1934}, which we approximate well using Equation~(\ref{eq:shevtar}) with $f=1$ and incorporating IA via Equation~(\ref{eq:lotz}) with \citet{Kaastra:AAS:1993} branching ratios. As with other very low charge states, Equation~(\ref{eq:shev5a}) can only match the Na$^{0+}$ cross section if one uses a much smaller value of $p_{0}$, here $p_{0} \sim 10$. 

Triple ionization of Al$^{2+}$ forming Al$^{5+}$ was measured by \citet{Steidl:1999}. These data can be well described by Equation~(\ref{eq:shevtar}) with $f=0.27$ from the triple ionization threshold of $\approx 300$~eV up to the peak of the cross section at $\approx 1000$~eV. However, the peak from the semiempirical formula is broader than that of the experimental result, resulting in a discrepancy of a factor of two at very high energies $\sim 5000$~eV. An alternative fitting scheme is to use Equation~(\ref{eq:shev5a}) with $p_{0}=28.6$, which matches the high energy data well, but rises too fast above threshold. Because we expect the low energy data to be the most important for typical applications, we have used the fit based on Equation~(\ref{eq:shevtar}). 

Considering our results for Al$^{2+}$ and for other isoelectronic sequences between Na-like through Ar-like, we estimate the triple and quadruple ionization cross section with Equation~(\ref{eq:shevtar}) setting $f=0.2$ and $f=1$, respectively. For quadruple ionization, this approximation, which is based on the DI energy threshold, probably underestimates the effective energy threshold due to the importance of indirect processes that turn on at higher energies. Quintuple and higher EIMI are neglected.

\subsection{Mg-like}\label{subsubsec:mg}

We estimate the double ionization cross sections of Mg-like ions using Equation~(\ref{eq:shev5a}) with $p_{0}=1$ and Equation~(\ref{eq:shev5b}). In order to match double ionization measurements of Mg$^{0+}$ \citep{McCallion:JPhysB:1992, Boivin:JPhysB:1998} and Al$^{1+}$ \citep{Steidl:1999} we need to increase $p_{0}$ in Equation~(\ref{eq:shev5a}) to $1.46$ and $2.24$, respectively, from the recommended value of $1$ \citep{Shevelko:JPhysB:2005}. However, these are small corrections to the overall cross sections, which are dominated by IA. This IA is probably from ionization of an $n=2$ electron, based on the apparent threshold. We model this indirect cross section for Mg-like ions, using Equation~(\ref{eq:shev5b}) and \citet{Kaastra:AAS:1993} branching ratios. This does, however, overestimate both the Mg$^{0+}$ and Al$^{1+}$ cross sections compared to experimental results by up to a factor of two. We found a similar, but smaller, discrepancy of about 30\% between the predicted double ionization cross section and the measurements for Ar$^{6+}$.

The reason for these discrepancies is not clear. Since the $n=2$ single ionization threshold is below the triple ionization threshold, an $n=2$ hole can relax either radiatively leading to a net single ionization or by ejecting a single electron leading to a net double ionization. Considering Mg$^{0+}$, if we were to ascribe the apparent excess branching ratio to single ionization rather than double ionization, the Mg$^{0+}$ single ionization cross section would increase by about  20\%. Similarly, for Al$^{1+}$ the increase to the single ionization cross section would be about 40\%. Such large contributions might appear in the single ionization data as an inflection in the cross section at the $n=2$ ionization threshold, however, such a features is not apparent in the single ionization data \citep{Belic:PRA:1987, McCallion:JPhysB:1992, Boivin:JPhysB:1998}. 

Triple ionization of Mg$^{0+}$ and Al$^{1+}$ can both be fit using Equation~(\ref{eq:shevtar}) with $f=1$. The fit matches the data for Mg$^{0+}$ well near threshold, but the peak cross section given by the fit is roughly half the experimental one. The fit for Al$^{1+}$ rises somewhat faster than the data near threshold, but the magnitude of the peak cross section is in excellent agreement. For both ions, these discrepancies may be due to additional MIMA processes. Quadruple ionization of Al$^{1+}$ can be described using Equation~(\ref{eq:shevtar}) with $f=0.27$.

We model triple and quadruple ionization of other Mg-like ions using Equation~(\ref{eq:shevtar}) with $f=0.2$ for triple ionization and $f=1$ for quadruple ionization. These values were chosen because they are consistent with the EIMI data for nearby isoelectronic sequences of Na-like through Ar-like ions.

\subsection{Al-like}\label{subsubsec:al}

Measurements of double ionization for Ar$^{5+}$ \citep{Tinschert:JPhysB:1989, Zhang:JPhysB:2002} are well described by Equations~(\ref{eq:shev5a}) and (\ref{eq:shev5b}) with the \citet{Shevelko:JPhysB:2005} parameters and \citet{Kaastra:AAS:1993} branching ratios, so we use these formulae for Al-like ions up to Ca ($Z \leq 20$). There is about 20\% discrepancy with the measurements for Ar$^{5+}$, which appears to be due to the IA being overestimated when using the branching ratios from \citet{Kaastra:AAS:1993}. One way to resolve this discrepancy would be to reduce the $2s$ branching ratio for double ionization from the \citet{Kaastra:AAS:1993} value of $\approx 0.99$ to $\approx 0.18$, where the excess branching ratio might go toward either single or triple ionization. Alternatively, the $2p$ branching ratio could be reduced, with the excess contributing to single ionization. Triple ionization measurements for Ar$^{5+}$ would be useful to test this. Since the resolution is ambiguous and the error is fairly modest, we still use the \citet{Kaastra:AAS:1993} branching ratios here.

For double ionization of Fe$^{13+}$, Equations~(\ref{eq:shev5a}) and (\ref{eq:shev5b}) become less accurate and tend to underestimate the direct ionization while overestimating the indirect contributions. We also find similar issues with later isoelectronic sequences. Those equations were developed by \citet{Shevelko:JPhysB:2005} for application to ``light positive ions'', so it is not surprising that they become less accurate for higher $Z$. Thus, for estimating the double ionization of Al-like ions with $Z \geq 21$ (above Sc), we use Equation~(\ref{eq:shevtar}) with $f=1$ and the other parameters from \citet{Belenger:JPhysB:1997} to model the DI cross section and add in IA contributions using Equation~(\ref{eq:lotz}).

We are not aware of any measurements of triple or higher order EIMI for Al-like ions. As mentioned above, we estimate the cross sections for these ions using Equation~(\ref{eq:shevtar}) with $f=0.2$ for triple ionization and $f=1$ for quadruple ionization. Again, this may underestimate the effective energy threshold, which is probably due to IA rather than the DI represented by Equation~(\ref{eq:shevtar}). 

\subsection{Si-like}\label{subsubsec:si}

\begin{figure}
	\centering \includegraphics[width=0.9\textwidth]{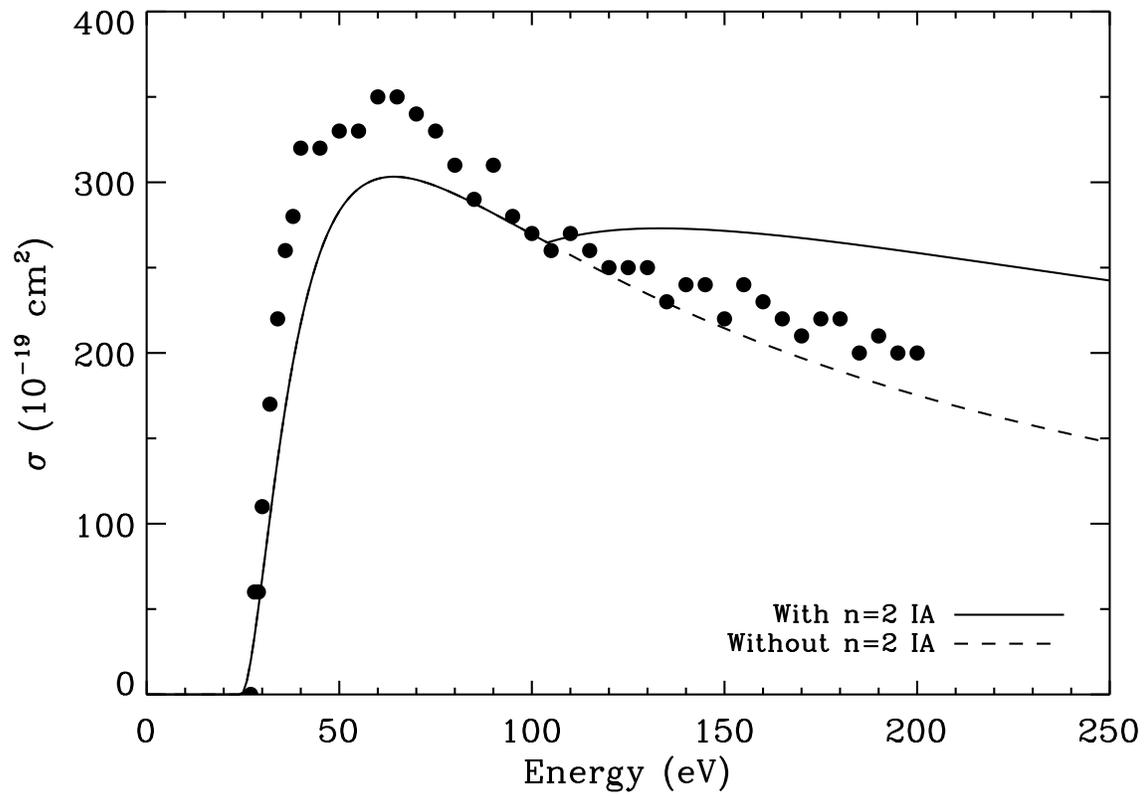}
	\caption{\label{fig:si02} Double ionization of Si$^{0+}$ forming Si$^{2+}$. Filled circles show the data of \citet{Freund:PRA:1990}. The solid curve shows our estimate using the branching ratios of \citet{Kaastra:AAS:1993}. The dashed curve illustrates the cross section of the case when the branching ratios are set to zero. It is clear that the true branching ratio is smaller than predicted.
	}
\end{figure}

The cross sections of Si-like systems can be estimated in the same way as those of Al-like ions. For $Z \leq 20$, double ionization is described by Equations~(\ref{eq:shev5a}) and (\ref{eq:shev5b}) with the parameters given by \citet{Shevelko:JPhysB:2005} for Ar$^{4+}$ and branching ratios from \citet{Kaastra:AAS:1993}. For $Z \geq 21$, we use Equation~(\ref{eq:shevtar}) with the parameters from \citet{Belenger:JPhysB:1997} and add IA using the Lotz cross section (Equation~\ref{eq:lotz}) and the \citet{Kaastra:AAS:1993} branching ratios. As was the case for Al-like ions, we find that the branching ratios for $2p$ IA given by \citet{Kaastra:AAS:1993} are probably overestimated. Figure~\ref{fig:si02} illustrates this for Si$^{0+}$, where the discrepancy with the measurements of \citet{Freund:PRA:1990} is particularly large. For this ion the $2s$ IA branching ratio predicted by \citet{Kaastra:AAS:1993} is already negligible, only $f_{\mathrm{BR}}\approx0.03$, and so the discrepancy must be due to the $2p$ branching ratio, which \citet{Kaastra:AAS:1993} predict $f_{\mathrm{BR}} \approx 1$. The triple and quadruple ionization cross sections are estimated following the same procedure as described for Al-like ions (Section~\ref{subsubsec:al}). 

\subsection{P-like}\label{subsubsec:p}

The total cross sections for double, triple, and quadruple ionization of P-like ions are estimated following the same procedures as for Si-like ions and using the appropriate parameters from \citet{Belenger:JPhysB:1997}, \citet{Shevelko:JPhysB:2005}, and \citet{Kaastra:AAS:1993} in the corresponding semiempirical formulae. We also now add quintuple ionization. Although there are no quintuple ionization data for P-like ions, we believe this is a reasonable place to start including quintuple ionization, because this is the first $M$-shell ion for which removing five electrons does not require breaking open the $L$-shell. We estimate the quintuple ionization cross sections based on Cl-like and Ar-like data and use Equation~(\ref{eq:shevtar}) with $f=0.5$ and the other parameters from \citet{Shevelko:JPhysB:1995}. We also include IA in the quintuple ionization cross section via Equation~(\ref{eq:lotz}) multiplied by the branching ratios, whenever they are predicted to be non-zero by \citet{Kaastra:AAS:1993}. 

\subsection{S-like}\label{subsubsec:s}

Double ionization measurements of S$^{0+}$ \citep{Ziegler:PSS:1982b, Freund:PRA:1990} and Ar$^{2+}$ \citep{Tinschert:JPhysB:1989} can be be fit using Equations~(\ref{eq:shev5a}) and (\ref{eq:shev5b}), but the $p_{0}$ parameter in Equation~(\ref{eq:shev5a}) needs to be increased from the value $p_{0}=40$ given by \citet{Shevelko:JPhysB:2005}. For sulfur, there is a discrepancy of about 30\% between the data of \citet{Ziegler:PSS:1982b} and those of \citet{Freund:PRA:1990}. For energies below the IA threshold, the \citeauthor{Ziegler:PSS:1982b} data can be fit by setting $p_{0} = 60$, while the \citeauthor{Freund:PRA:1990} measurements are consistent with $p_{0}=40$. For Ar$^{2+}$, the experiments of \citet{Tinschert:JPhysB:1989} are best fit with a value of $p_{0}\approx56$. Based on these results, $p_{0}\approx60$ is more consistent with the available data and so we use this value for estimating the direct double ionization cross section for other S-like ions that have $Z \leq 20$. 

For double ionization of $Z \geq 21$ ions, we follow the same procedure as for the previous isoelectronic sequences and use Equations~(\ref{eq:shevtar}) and (\ref{eq:lotz}). One representative cross section from this group exists, which is a measurement of double ionization for Ti$^{6+}$ by \citet{Hartenfeller:JPhysB:1998}. For that particular ion, we use the fits given by \citet{Shevelko:JPhysB:2006}, Equations~(\ref{eq:shev6a}) and (\ref{eq:shev6b}). For comparison with the $Z \leq 20$ ions, we have found that a value of $p_{0}=180$ would be needed in  Equation~(\ref{eq:shev5a}) in order to approximately match the magnitude of the direct ionization cross section of $Z \geq 21$ ions. This is much larger than the value of $p_{0} = 60$ of the low $Z$ data and illustrates that the \citet{Shevelko:JPhysB:2005} formulae are less predictive for heavy ions. 

\begin{figure}
	\centering \includegraphics[width=0.9\textwidth]{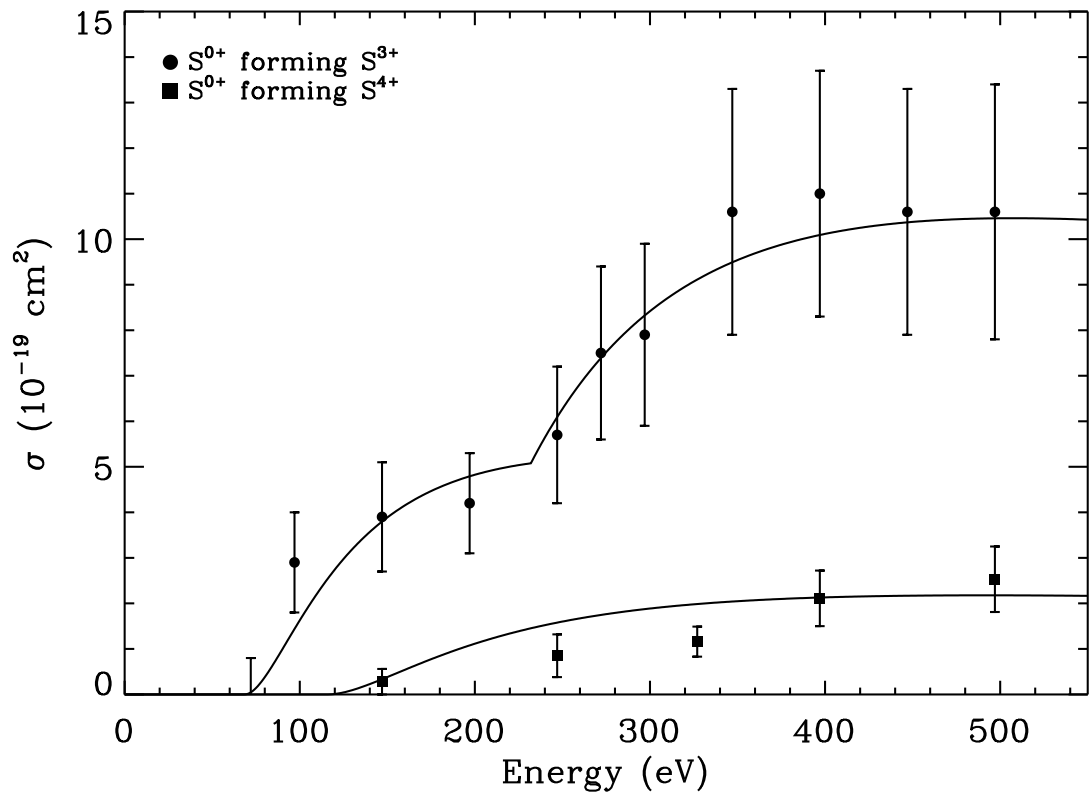}
	\caption{\label{fig:s03} Triple ionization of S$^{0+}$ forming S$^{3+}$ (filled circles) and quadruple ionization of S$^{0+}$ forming S$^{4+}$ (filled squares), from the measurements of \citet{Ziegler:PSS:1982b}. The solid curves show our fits to the data using the semiempirical formulae, as discussed in Section~\ref{subsubsec:s}. 
	}
\end{figure}

Triple ionization has been measured for S$^{0+}$ \citep{Ziegler:PSS:1982b} and Ar$^{2+}$ \citep{Muller:PRL:1980}. Both of these cross sections can be accurately represented using Equations~(\ref{eq:shevtar}) and (\ref{eq:lotz}). For S$^{0+}$ the best fit has $f=0.29$ in Equation~(\ref{eq:shevtar}) and for Ar$^{2+}$ $f=0.18$. The other parameters are given by \citet{Shevelko:JPhysB:1995} with IA branching ratio data from \citet{Kaastra:AAS:1993}. The S$^{0+}$ cross section is illustrated in Figure~\ref{fig:s03}. Based on these values and those for nearby isoelectronic sequences, we model triple ionization of other S-like ions using $f=0.2$ in Equation~(\ref{eq:shevtar}), as is done for other Na-like through Zn-like ions. 

The quadruple ionization of S$^{0+}$ has been measured by \citet{Ziegler:PSS:1982b} and is well-described by Equation~(\ref{eq:shevtar}) with $f=1$ (Figure~\ref{fig:s03}). We also include IA using the Lotz cross section and \citet{Kaastra:AAS:1993} branching ratios, although the relevant IA threshold is at much higher energies than were included in the measurements. We  estimate the cross sections for quadruple ionization of other S-like ions using the same scheme as for S$^{0+}$. Quintuple ionization is included following the procedure described in Section~\ref{subsubsec:p}. 

\subsection{Cl-like}\label{subsubsec:cl}

An extensive set of measurements covering double through septuple ionization of the Cl-like ion Ar$^{1+}$ has been reported by \citet{Belic:JPhysB:2010}. These data are complemented by the double and triple ionization measurements of \citet{Muller:PRL:1980} and \citet{Muller:JPhysB:1985}. We have estimated empirical branching ratios for IA based on this extensive dataset, which gives us an estimate of the possible errors in \citet{Kaastra:AAS:1993} data for $L$-shell vacancies.  Our estimates are limited to the $L$-shell, because the data do not extend to high enough energies to see the $K$-shell IA. The IA contributions are most obvious in the higher order EIMI cross sections. For this reason, we discuss comparison to the Ar$^{1+}$ data going backwards from septuple ionization. Table~\ref{table:far1} summarizes our inferred IA branching ratios for IA of Ar$^{1+}$ and compares them to the original values from \citet{Kaastra:AAS:1993}.

The experimental data for septuple, sextuple, and quintuple ionization of Ar$^{1+}$ forming Ar$^{8+}$, Ar$^{7+}$, and Ar$^{6+}$ all have thresholds that appear to be several hundred eV higher than the direct ionization thresholds. This suggests that the ionization mechanism is dominated by an indirect process. However, the septuple and sextuple ionization thresholds are above the single ionization threshold for an $L$-shell electron, which indicates the indirect process is not IA from the creation of an $L$-shell hole. In the case of quintuple ionization, an $L$-shell vacancy could energetically decay by emission of four additional electrons, but since the experimental data show no cross section until well above the $L$-shell ionization threshold, the branching ratio for that process must be small. The relevant indirect process for these EIMI cross sections may be MIMA, where multiple direct ionization of core electrons are followed by the ejection of multiple electrons as the resulting excited state relaxes (see e.g., Section~\ref{subsubsec:n}). For these complex ions, there are several combinations of inner-shell vacancies that could match the apparent threshold, so it is not possible to empirically estimate a model cross section based on MIMA. Instead, we approximately match the magnitude of the measurements using Equation~(\ref{eq:shevtar}) with $f=0.055$ for septuple ionization, $0.24$ for sextuple ionization, and $0.52$ for quintuple ionization, though this scheme underestimates the effective threshold energy. In each case, $K$-shell IA is accounted for via Equation~(\ref{eq:lotz}) multiplied by the branching ratios from \citet{Kaastra:AAS:1993}. 

\begin{figure}
	\centering \includegraphics[width=0.9\textwidth]{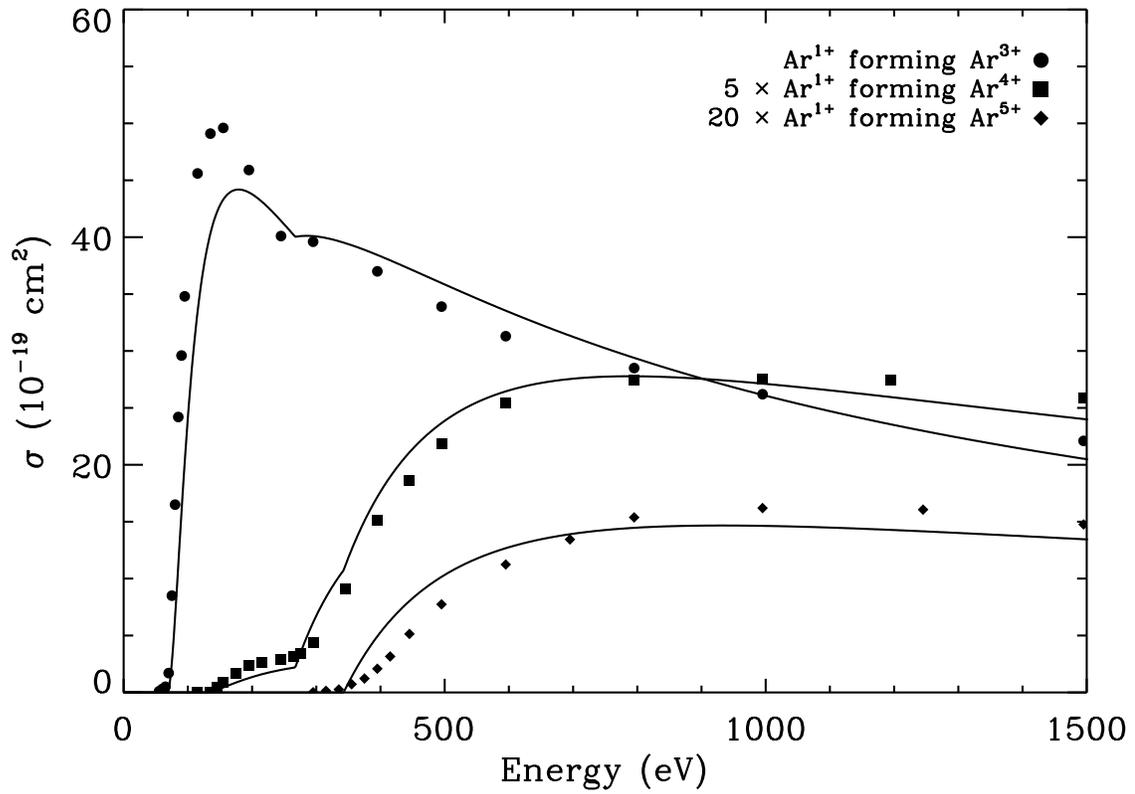}
	\caption{\label{fig:ar1345} Double, Triple, and Quadruple ionization of Ar$^{1+}$ forming Ar$^{3+}$, Ar$^{4+}$, and Ar$^{5+}$, respectively. The symbols show the data of \citet{Belic:JPhysB:2010}. For readability, the triple ionization data have been multiplied by a factor of $5$ and the quadruple ionization data by a factor of $20$. Our cross sections using the semiempirical formulae are illustrated by the solid curves. 
	}
\end{figure}

The observed energy threshold for quadruple ionization matches very well with the threshold for single ionization of a $2s$ electron, of $\approx 343$~eV \citep{Shevelko:JPhysB:2005}. Since the data show no cross section between the direct ionization threshold of $\approx 203$~eV and the $2s$-ionization threshold, direct ionization and $2p$ IA are negligible. The observed cross section can be fit using the Lotz cross section (Equation~\ref{eq:lotz}) if the branching ratio is $f_{\mathrm{BR}}=0.26$. This is in contrast to \citet{Kaastra:AAS:1993}, who predict the branching ratio to be $f_{\mathrm{BR}}=0$. Our empirical estimate assumes that the entire cross section is due to IA from a $2s$ hole, but in reality other indirect processes may also contribute. Figure~\ref{fig:ar1345} compares our cross section to the measurements of \citet{Belic:JPhysB:2010}. 

Figure~\ref{fig:ar1345} also illustrates the triple ionization cross section, which does show evidence for direct ionization that can be well matched by Equation~(\ref{eq:shevtar}) with $f=0.095$ plus IA added via the Lotz formula multiplied by branching ratios adjusted as follows: The cross section shows a rapid increase below $300$~eV, which corresponds well with the $2p$ single ionization threshold. In order to infer a branching ratio for $2p$ IA, we first reduce the \citeauthor{Kaastra:AAS:1993} branching ratio for $2s$ IA leading to triple ionization by $0.26$, because we believe that fraction goes to quadruple ionization. This leaves a remaining branching ratio of $f_{\mathrm{BR}}=0.7048$ to go toward $2s$ IA triple ionization. Once that is fixed, a least squares fit to the experimental data gives the $f_{\mathrm{BR}}=0.222$ for the $2p$ IA triple ionization. For comparison, \citet{Kaastra:AAS:1993} predicts that the branching ratio for a $2p$ hole to lead to triple ionization is zero. Here, there is also additional uncertainty because there is a large discrepancy between the triple ionization measurements, with the cross section of \citet{Muller:PRL:1980} being about 40\% smaller than the results of \citet{Belic:JPhysB:2010}. For consistency, we have been using the \citeauthor{Belic:JPhysB:2010} data throughout this section.

For double ionization, there is good agreement between the \citet{Muller:JPhysB:1985} and \citet{Belic:JPhysB:2010} data. In order to describe Ar$^{1+}$ double ionization, we adjust the branching ratios from \citet{Kaastra:AAS:1993} downwards to account for the branching ratios that we empirically ascribe to triple and quadruple ionization. We then fit the data using Equations~(\ref{eq:shev5a}) and (\ref{eq:shev5b}) in order to determine the best value for $p_{0}$ in Equation~(\ref{eq:shev5a}), which we find to be $p_{0}=85.5$. The resulting cross section is compared to the experiments of \citet{Belic:JPhysB:2010} in Figure~\ref{fig:ar1345}. This value of $p_{0}$ disagrees with the one given by \citet{Shevelko:JPhysB:2005} who found $p_{0}=67$, which  was also based on Ar$^{1+}$ measurements. The discrepancy may be due to our reducing the branching ratios for indirect processes in the Ar$^{1+}$ double ionization cross section. A value of $p_{0}=67$ does seem to be a good fit to double ionization data for Cl$^{0+}$ \citep{Freund:PRA:1990}. To fit the data for Ti$^{5+}$ or Fe$^{9+}$, we would require larger values of $p_{0}$, but this is consistent with other isoelectronic sequences where we have found that Equations~(\ref{eq:shev5a}) and (\ref{eq:shev5b}) become less predictive for $Z \geq 21$ ions. Instead we can match the Ti$^{5+}$ data using Equations~(\ref{eq:shev6a}) and (\ref{eq:shev6b}) and the fits given by \citet{Shevelko:JPhysB:2006}. The Fe$^{9+}$ measurements are well described using Equation~(\ref{eq:shevtar}) with $f=1$ plus Equation~(\ref{eq:lotz}) for the IA contributions.

Based on the above results, we extrapolate the EIMI cross sections for other Cl-like ions in the following way: Double ionization of ions with $Z \leq 20$ is described using Equations~(\ref{eq:shev5a}) and (\ref{eq:shev5b}) with the parameters given in \citet{Shevelko:JPhysB:2005}. Double ionization of ions with $Z \geq 21$ uses Equation~(\ref{eq:shevtar}) with $f=1$ and Equation~(\ref{eq:lotz}). For triple ionization, we use Equation~(\ref{eq:shevtar}) with $f=0.2$, as it is for all sequences from Na-like through Zn-like, and incorporate IA using Equation~(\ref{eq:lotz}). Quadruple ionization is predicted by Equation~(\ref{eq:shevtar}) with $f=1$ (as for all other Na-like through Ar-like isoelectronic sequences) and IA via the Lotz formula. For quintuple, sextuple, and septuple ionization we also use Equation~(\ref{eq:shevtar}) with $f=0.5$ for quintuple, $f=0.2$ for sextuple, and $f=0.07$ for septuple ionization. Since there is very little data on these processes, we apply the same scheme to all quintuple, sextuple, and septuple EIMI of isoelectronic sequences Cl-like through Zn-like. For all of the above cases, except for argon, we use the \citet{Kaastra:AAS:1993} branching ratios.  

\subsection{Ar-like}\label{subsubsec:ar}

There is an extensive set of measurements for EIMI for Ar$^{0+}$ from double to septuple ionization \citep{Schram:Physica:1966b, Krishnakumar:JPhysB:1988, Syage:JPhysB:1991, McCallion:JPhysB:1992a, Almeida:JESRP:1994, Straub:PRA:1995, Rejoub:PRA:2002}. As with the Cl-like ions above, we consider the data going backwards from septuple ionization and make empirical estimates for the indirect ionization branching ratios. Table~\ref{table:far0} summarizes the inferred IA branching ratios for IA of Ar$^{0+}$ and compares them to the original values from \citet{Kaastra:AAS:1993}.

Direct septuple ionization of Ar$^{0+}$ forming Ar$^{7+}$ can be described by Equation~(\ref{eq:shevtar}) with $f=0.094$ at energies below the $K$-shell ionization threshold of about $3200$~eV. For higher energies, we assume that the main indirect process is $K$-shell single ionization followed by sextuple autoionization and that this can be represented using the Lotz formula multiplied by a branching ratio. A fit to the data of \citet{Schram:Physica:1966b} implies a branching ratio of $f_{\mathrm{BR}}=0.014$. For comparison, \citet{Kaastra:AAS:1993} predict no branching ratio for a net septuple ionization following a $K$-shell vacancy. Also, note that single ionization of an $L$-shell electron would not create a state with sufficient internal excitation to decay in a way that leads to septuple ionization, however, multiple ionization or excitation of the $L$-shell might contribute and we have not accounted for those effects.

Sextuple ionization can be described similarly, using Equation~(\ref{eq:shevtar}) with $f=0.22$ and adding $K$-shell IA via the Lotz cross section with the branching ratio $f_{\mathrm{BR}}=0.0544$, from \citet{Kaastra:AAS:1993}. Here, the \citet{Kaastra:AAS:1993} branching ratio appears reasonable. One issue with the sextuple ionization data is that there is a large discrepancy between the measurements, with the cross section of \citet{Schram:Physica:1966b} a factor of two larger than that of \citet{Almeida:JESRP:1994} below $2000$~eV, which is the maximum energy measured by \citet{Almeida:JESRP:1994}. Our fit lies in between these two measurements below $2000$~eV, but matches \citet{Schram:Physica:1966b} results fairly well at higher energies.

The experimental quintuple ionization threshold appears to be at least 100~eV above the theoretical DI threshold. The threshold corresponds well, instead, with that for single ionization of a $2s$ electron at $\approx 327$~eV. This suggests that DI is negligible, although the measurements near the DI threshold are sparse. By fitting Equation~(\ref{eq:lotz}) to the data, we estimate the IA branching ratio from $2s$ IA to be about $f_{\mathrm{BR}}=0.038$, but given the spread in the experimental measurements the uncertainty is about a factor of two. We also add in $K$-shell IA using the Lotz cross section and the \citet{Kaastra:AAS:1993} branching ratio adjusted down to $0.1869$ to account for the branching ratio we ascribed to septuple ionization. Since we would like to estimate the cross sections for other systems that lack experimental data, it is useful to also fit these data using Equation~(\ref{eq:shevtar}). Doing so, we find that the magnitude of the data would be well-described with $f\approx0.4$, although the threshold corresponds to the direct threshold and so is too low. 

The measured quadruple ionization threshold lies at $\sim 300$~eV, well above the theoretical direct ionization threshold of about $144$~eV. Energetically, a $2p$ vacancy could lead to quadruple ionization, but the threshold matches better with the higher energy threshold for single ionization of the $2s$ subshell. A fit to the data using the Lotz cross section finds that the $2s$ IA branching ratio for quadruple ionization is $f_{\mathrm{BR}}=0.46$. To fully describe the cross section we also add $K$-shell IA via the Lotz cross section multiplied by the \citet{Kaastra:AAS:1993} branching ratio. Alternatively, the magnitude of the cross section could be approximately matched using Equation~(\ref{eq:shevtar}) with $f=1$, but the threshold would be too low compared to experiment. 

The direct triple ionization of Ar$^{0+}$ forming Ar$^{3+}$ exhibits a narrow peak that can best be fit using Equation~(\ref{eq:shev5a}) with $p_{0}=22.2$. For the indirect contributions, we use Equation~(\ref{eq:shev5b}). The branching ratios for $2s$ IA are adjusted down by about half from the \citet{Kaastra:AAS:1993} branching ratios in order to account for the branching ratio we ascribe to higher order EIMI. Then we fit the data of \citet{Almeida:JESRP:1994} to find a $2p$ branching ratio of $\approx 0.25$. The scatter in the various measured cross sections is roughly $20\%$. 

\begin{figure}
	\centering \includegraphics[width=0.9\textwidth]{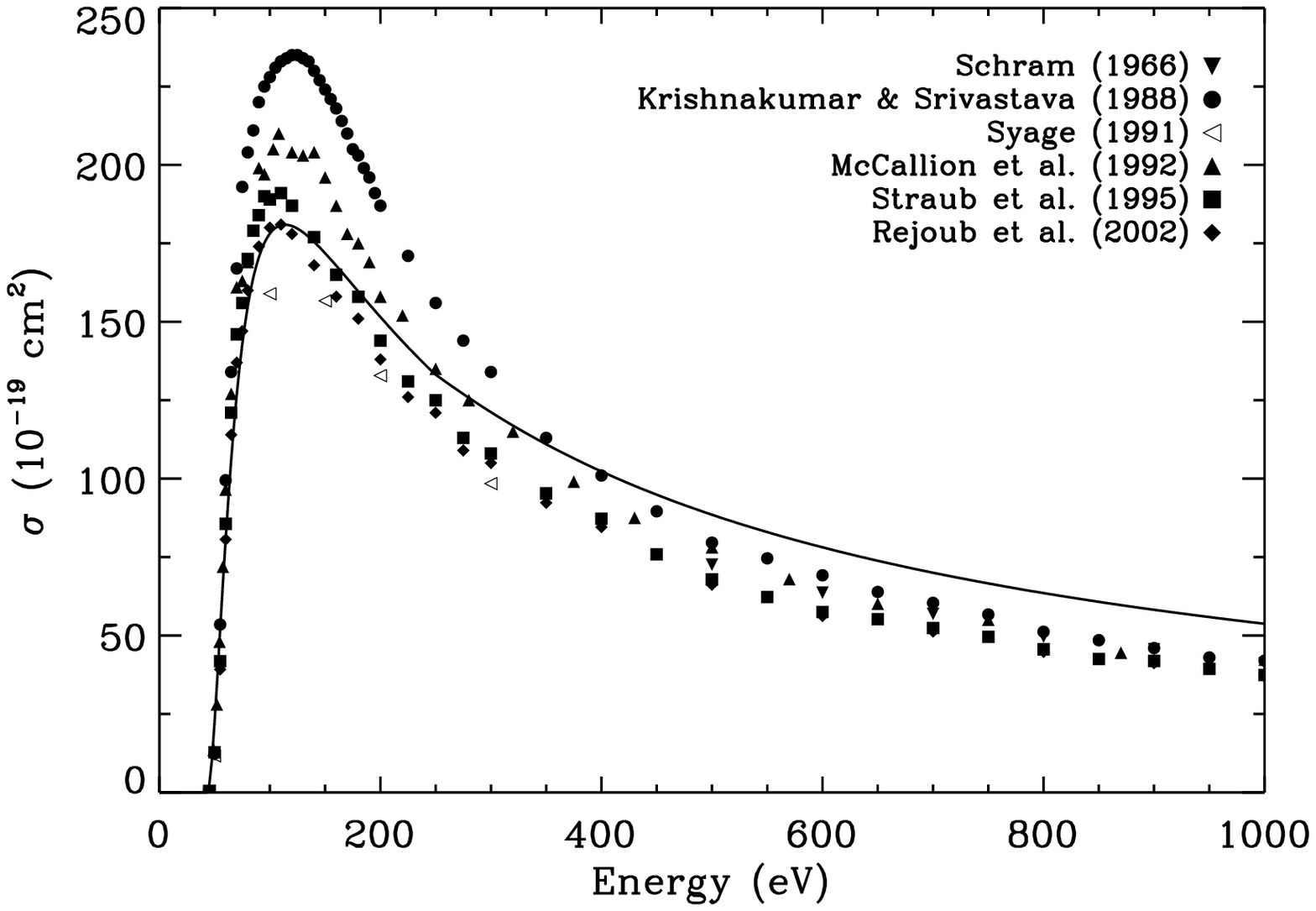}
	\caption{\label{fig:ar02} Double ionization of Ar$^{0+}$ forming Ar$^{2+}$. The symbols show the data of \citet[downward-pointing filled triangles]{Schram:Physica:1966b}, \citet[][filled circles]{Krishnakumar:JPhysB:1988}, \citet[][open triangles]{Syage:JPhysB:1991}, \citet[][filled triangles]{McCallion:JPhysB:1992a}, \citet[][filled squares]{Straub:PRA:1995}, and \citet[][filled diamonds]{Rejoub:PRA:2002}. Our fit, described in the text, is illustrated by the solid curve.
	}
\end{figure}

Double ionization of Ar$^{0+}$ is described using Equations~(\ref{eq:shev5a}) and (\ref{eq:shev5b}). For the direct cross section we find a best fit with $p_{0}=89.5$ in Equation~(\ref{eq:shev5a}). This is somewhat larger than the value of $p_{0}=80.0$ given by \citet{Shevelko:JPhysB:2005}. The reason for the discrepancy seems to be that \citeauthor{Shevelko:JPhysB:2005} based their fit on the data of \citet{Syage:JPhysB:1991}, which are systematically low compared to the various other measurements (see Figure~\ref{fig:ar02}). For the indirect cross sections, included via Equation~(\ref{eq:shev5b}), we use the \citet{Kaastra:AAS:1993} branching ratios, with the $2p$ branching ratio reduced by $0.25$ to account for the branching ratio we ascribed to triple ionization. 

For double ionization of other Ar-like ions with $Z \leq 20$, we follow the same procedure and use Equation~(\ref{eq:shev5a}) with $p_{0}=89.5$ and Equation~(\ref{eq:shev5b}), but with the unmodified \citet{Kaastra:AAS:1993} branching ratios. Comparing this scheme to the K$^{1+}$ double EIMI data of \citet{Hirayama:JPSJ:1986}, we find that the direct ionization part (Equation~\ref{eq:shev5a}) matches quite well. However, we overestimate IA, which is probably because the \citet{Kaastra:AAS:1993} predictions tend to underestimate the branching ratios for ejecting more electrons, and therefore overestimate the branching ratios for ejecting fewer electrons (see Table~\ref{table:far0}). For double ionization of heavier ions, $Z \geq 21$, we estimate the cross sections with Equation~(\ref{eq:shevtar}) with $f=1$ and Equation~(\ref{eq:lotz}) with the branching ratios from \citet{Kaastra:AAS:1993}. 

The direct ionization of other Ar-like ions is modeled by Equation~(\ref{eq:shevtar}) with $f=0.2$ for triple ionization, $f=1$ for quadruple ionization, $f=0.5$ for quintuple ionization, $f=0.2$ for sextuple ionization, and $f=0.07$ for septuple ionization. For each of these ions, we also add IA contributions via the Lotz cross section and the \citet{Kaastra:AAS:1993} branching ratios. We do not extrapolate our scaled branching ratios, because there is not enough data available to understand the variation of $f_{\mathrm{BR}}$ with $Z$ within the isoelectronic sequence. In some cases, such as for quadruple and quintuple ionization, this scheme represents the cross section as direct ionization when in fact it is probably dominated by IA, as it is for Ar$^{0+}$. However, we lack a good method for extrapolating branching ratios. The above procedure is expected to at least estimate the magnitude of the cross sections, but the practical threshold is likely higher in energy than the DI threshold.

\subsection{K-like through Zn-like}\label{subsubsec:kzn}

For ions in the K-like through Zn-like isoelectronic sequences, there are few data available to compare with. For this reason, we generally estimate the cross sections for ions in these isoelectronic sequences all following the same procedure. The exception is for those cases where there are experimental data, where we try to use the most accurate formulae even if they are not necessarily useful for predicting the behavior of other cross sections. 

\citet{Shevelko:JPhysB:2006} gives fits to experimental double ionization cross sections for low charge states of Ti$^{1 - 6+}$, Fe$^{1, 3 - 6+}$, and Ni$^{1-6+}$ using Equations~(\ref{eq:shev6a}) and (\ref{eq:shev6b}). For these ions, we use the \citet{Shevelko:JPhysB:2006} fit parameters. We also use the same formulae to describe the Fe$^{0+}$ data of \citet{Shah:JPhysB:1993} with the fit parameters we found earlier \citep{Hahn:ApJ:2015}. 

Measurements also exist for double ionization of Sc$^{1+}$ \citep{Jacobi:JPhysB:2005} and Cu$^{0+}$ \citep{Freund:PRA:1990, Bolorizadeh:JPhysB:1994}. \citet{Jacobi:JPhysB:2005} found that Sc$^{1+}$ measurements were well described using Equation~(\ref{eq:shevtar}) for the direct cross section and adding to IA from ionization of a $3p$ electron using the Lotz cross section (Equation~\ref{eq:lotz}) with $f_{\mathrm{BR}}=0.68$. They find no contribution from the $3s$-IA. These inferred branching ratios are well below the predictions of \citet{Kaastra:AAS:1993} who predict $f_{\mathrm{BR}}=1$ for IA arising from both $3s$ and $3p$ subshells. We find that the Cu data of \citet{Freund:PRA:1990} and \citet{Bolorizadeh:JPhysB:1994} can best be fit using Equation~(\ref{eq:shev5a}) with $p_{0}=10.44$ and Equation~(\ref{eq:shev5b}) with the \citet{Kaastra:AAS:1993} branching ratios for IA. 

In order to predict the double ionization cross sections of all the unmeasured K-like through Zn-like systems, we take the predictions of using Equation~(\ref{eq:shevtar}) with $f=1$ and adding IA contributions via the Lotz formula and compare to the cross sections reported by \citet{Shevelko:JPhysB:2006}. This method gives a reasonable alternative fit to these data, and so we used this procedure to estimate the cross sections for the unmeasured systems.

Triple ionization of Ti ions have been reported by \citet{Hartenfeller:JPhysB:1998}. We found that these triple ionization cross sections could be described using Equation~(\ref{eq:shevtar}) plus IA through the Lotz cross section. In Equation~(\ref{eq:shevtar}) we find that for triple ionization, $f=0.1$ fits K-like Ti$^{3+}$, $f=0.2$ fits Ca-like Ti$^{2+}$ and $f=0.25$ fits Sc-like Ti$^{1+}$. 

\citet{Jacobi:JPhysB:2005} measured triple ionization of Sc$^{1+}$ and found that the magnitude of the cross section was approximated well by Equation~(\ref{eq:shevtar}), although the peak of the model cross section is broader than what was measured. \citeauthor{Jacobi:JPhysB:2005} also inferred an $2p$ IA branching ratio of $f_{\mathrm{BR}}\approx 0.9$, which is in rough agreement with \citet{Kaastra:AAS:1993}. Based on their results, we represent the total Sc$^{1+}$ triple ionization cross section as a sum of the direct contribution using Equation~(\ref{eq:shevtar}) with $f=1$ plus the indirect cross section using Equation~(\ref{eq:lotz}) with branching ratios from \citet{Kaastra:AAS:1993}. This matches the measurements well near the threshold, but above about $400$~eV the cross section is overestimated by about 50\%. 

The Cu$^{0+}$ triple ionization measurements of \citet{Bolorizadeh:JPhysB:1994} can be fit using the same scheme, setting $f=0.16$ in Equation~(\ref{eq:shevtar}). In all cases the IA branching ratios are from \citet{Kaastra:AAS:1993}. Based on these results, we estimate the triple ionization cross sections for all other K-like through Zn-like ions using the same procedure and setting $f=0.2$ in Equation~(\ref{eq:shevtar}). 

Quadruple ionization can also be described using Equation~(\ref{eq:shevtar}) plus the indirect contributions using Equation~(\ref{eq:lotz}) with the branching ratios from \citet{Kaastra:AAS:1993}. For Ti$^{1+}$ \citep{Hartenfeller:JPhysB:1998} we find $f=0.25$ in Equation~(\ref{eq:shevtar}), while for Cu$^{0+}$ \citep{Bolorizadeh:JPhysB:1994} $f=0.1$. \citet{Jacobi:JPhysB:2005} found that their quadruple ionization measurements of Sc$^{1+}$ could be matched by using $f=0.15$ in Equation~(\ref{eq:shevtar}) and branching ratios of $f_{\mathrm{BR}}=0.02$ for $2p$ IA and $f_{\mathrm{BR}}=0.8$ for $2s$ IA. Since their estimated branching ratios are very close to the predictions of \citet{Kaastra:AAS:1993}, we continue to use the \citeauthor{Kaastra:AAS:1993} results in our compilation.

For ions that have not been measured, we err on the conservative end of the estimated $f$ values for Equation~(\ref{eq:shevtar}) and set $f=0.1$ in estimating the direct quadruple ionization cross sections for other ions in these isoelectronic sequences. We model the indirect cross sections as IA using the Lotz cross section and \citet{Kaastra:AAS:1993} branching ratios.

Quintuple ionization measurements for these isoelectronic sequences have been performed only for Sc$^{1+}$ \citep{Jacobi:JPhysB:2005} and Cu$^{0+}$ \citep{Bolorizadeh:JPhysB:1994}. For Sc$^{1+}$, \citet{Jacobi:JPhysB:2005} found that the measurements could be matched up to about 590~eV using Equation~(\ref{eq:shevtar}) with $f=0.008$ plus IA represented by the Lotz cross section with $f_{\mathrm{BR}}=0.0006$ for the $2p$ subshell and $0.055$ for the $2s$ subshell. Above 590~eV they found additional indirect ionization cross sections, which they ascribed to a MIMA process involving direct double ionization of inner shell electrons followed by autoionization of three additional electrons. Since it is difficult to model these complex ionization processes given the limited available data, we represent the Sc$^{1+}$ quintuple ionization cross section using Equation~(\ref{eq:shevtar}) with $f=0.5$, which matches the peak of the experimental data, but overestimates the cross section between the DI threshold at $\approx 313$~eV and the onset of the unknown ionization processes at 590~eV.

For Cu$^{0+}$, the quintuple ionization cross section is well described by Equation~(\ref{eq:shevtar}) with $f=0.65$ and adding in IA contributions using Equation~(\ref{eq:lotz}) multiplied by the branching ratios from \citet{Kaastra:AAS:1993}. Based on this, and also considering our quintuple ionization results for Ar-like and Cl-like ions, we estimate all other quintuple ionization cross sections using the same scheme, but with $f=0.5$ in Equation~(\ref{eq:shevtar}). 

As far as we know, there are no sextuple or septuple ionization measurements for K-like through Zn-like ions. Based on our results for Ar-like and Cl-like ions, we estimate these cross sections using Equations~(\ref{eq:shevtar}) and Equation~(\ref{eq:lotz}) multiplied by the IA branching ratios. For Equation~(\ref{eq:shevtar}) we set $f=0.2$ for sextuple ionization and $f=0.07$ for septuple ionization. 

In some cases \citet{Kaastra:AAS:1993} predict branching ratios for higher order EIMI beyond septuple ionization. In those cases, we only use Equation~(\ref{eq:lotz}) multiplied by the \citeauthor{Kaastra:AAS:1993} branching ratios. Because DI contributions are likely to be small and there are no data, we ignore DI for EIMI beyond septuple ionization.


\subsection{Uncertainties}\label{subsubsec:unc}
	
Given the sparseness of experimental EIMI data with which to compare, it is difficult to quantitatively describe their uncertainty. Nevertheless, we can give at least some rough estimate for the accuracy of our interpolations. For double ionization, there is data for most isoelectronic sequences and the semiempirical formulae seem to perform quite well. On this basis, we would estimate our double ionization cross sections to be accurate to $\sim30\%$. There is significantly less data available for triple ionization, and our cross sections are probably accurate to about a factor of two or better. For higher order EIMI our estimates are essentially guesses and we can only say that they are probably within an order of magnitude. 

For many quadruple and higher order EIMI cross sections, we have estimated the cross sections using Equation~(\ref{eq:shevtar}), which is a scaling for DI. These estimates are adequate for matching the magnitude of the cross sections in the limited available data. However, indirect processes are probably more important than DI and this scheme probably understates the effective energy threshold. For plasmas where the electron distribution includes many electrons below the true effective energy threshold, our estimates may have the effect of generating too high of a rate coefficient for the corresponding ionization process. 
	
\section{Summary and Future Needs}\label{sec:conclusions}

We report the first comprehensive set of EIMI cross sections for the astrophysically relevant systems, He-like through Zn-like. These data will be useful for the modeling of collisionally ionized plasmas that are subject to rapid heating or that have non-thermal electron distributions. Up to now, the lack of EIMI data has led most researchers to ignore EIMI. Our data can be used to quantify the effect of EIMI in the modeling of astrophysical systems. The application of these kinds of data can be expected to grow as the precision of astrophysical diagnostics requires increasingly precise atomic data for interpretation. 

In reviewing the data we have used semiempirical formulae and have identified some of their limitations. Equation~(\ref{eq:shevtar}) tends to overestimate the EIMI cross section and so we have introduced a scaling factor $f < 1$ in order to match the experimental data. We often resort to this formula, for describing triple and higher order EIMI because it appears to predict the magnitude well, although it does not account for IA processes, which can dominate those cross sections. As a result, this formula can significantly underestimate the effective EIMI threshold which is often higher than the DI threshold. 

In some cases also, the shape of the cross section predicted by Equation~(\ref{eq:shevtar}) does not match the data very well, particularly near the peak cross section. For example, the sharp peaks in the direct triple ionization data for N$^{1+}$, Ne$^{0+}$, and Ar$^{0+}$ are better matched by a fit based on Equation~(\ref{eq:shev5a}), which was developed for direct double ionization. It appears that this problem is prevalent for near-neutrals, but this may be only an experimental bias since most of the existing triple ionization experiments are for neutrals and singly charged ions. 

We found that Equation~(\ref{eq:shev5a}) works very well for describing direct double ionization of light ions ($Z \leq 20$), though with some caveats. For near-neutrals we have found that it is usually necessary to reduce the $p_{0}$ value from that given by \citet{Shevelko:JPhysB:2005}. For heavier ions ($Z \geq 21$), the parameter $p_{0}$ varies significantly within the isoelectronic sequence so that the formula is no longer predictive, and we use instead Equation~(\ref{eq:shevtar}) with the parameters of \citet{Belenger:JPhysB:1997}. 

IA is included in our cross sections by using either Equation~(\ref{eq:shev5b}) or (\ref{eq:lotz}). Generally we use Equation~(\ref{eq:shev5b}) when DI is described by Equation~(\ref{eq:shev5a}) and we use the Lotz formula (Equation~\ref{eq:lotz}) when DI is described by Equation~(\ref{eq:shevtar}). In either case the cross sections are scaled by the branching ratios $f_{\mathrm{BR}}$, for which we have relied mainly on \citet{Kaastra:AAS:1993}. For some cases, we were able to infer $f_{\mathrm{BR}}$ from the data and compare to the predictions of \citeauthor{Kaastra:AAS:1993}.  Our comparison shows that \citeauthor{Kaastra:AAS:1993} tend to underestimate the branching ratios for ejecting many electrons and so overestimate the branching ratios for ejecting fewer electrons (see Tables~\ref{table:far1} and \ref{table:far0}). 

A systematic factor that may contribute to the discrepancy between the theoretical and experimental IA branching ratios is the effect of excitation during the initial collision. \citet{Kucas:ApJ:2015} calculated autoionization branching ratios for $K$-shell vacancies for various charge states of Ne, Mg, Si, S, and Ar. In their calculations they included the ground configuration with a $K$-shell vacancy as well as excited configurations that had both a $K$-shell vacancy plus additional excitation. For the ground configurations, the \citet{Kucas:ApJ:2015} branching ratios are very similar to the results of \citet{Kaastra:AAS:1993}. However, when the initial configuration is more excited, there are substantial branching ratios for the ejection of additional electrons. Thus, if an electron-ion collision leads to excitation in addition to the ionization of a core electron, the branching ratio for higher order EIMI processes can be higher than predicted based on the \citeauthor{Kaastra:AAS:1993} branching ratios. Unfortunately, it is not feasible to perform a detailed comparison of the \citet{Kucas:ApJ:2015} predictions with the available EIMI cross sections. This is because for the ions that have branching ratio calculations, the EIMI measurements are most detailed at low energies, below the $K$-shell IA threshold, and because the cross sections are dominated by $L$-shell IA so that $K$-shell IA is small and not well resolved.

Because of the limited available experimental data and the lack of reliable theory, we expect that these estimated cross sections will need to be revised in the future. Currently, double ionization measurements are the most comprehensive, but even for double ionization there are few measurements for high charge states or for heavy ions. For higher order EIMI the dearth of experimental data is much worse. For many isoelectronic sequences, such as Be-like, O-like, Al-like, Si-like, and P-like, we could not find any triple ionization measurements. Most isoelectronic sequences lack any measurements of quadruple or higher order EIMI. 

The lower order EIMI cross sections are the largest, with the magnitude of the cross sections typically falling by an order of magnitude for each additional ejected electron. Thus, the lower order EIMI cross sections are expected to be the most important and it would make sense to prioritize measurements of double and triple ionization to fill in the existing gaps. 

\begin{figure}
	\centering \includegraphics[width=0.9\textwidth]{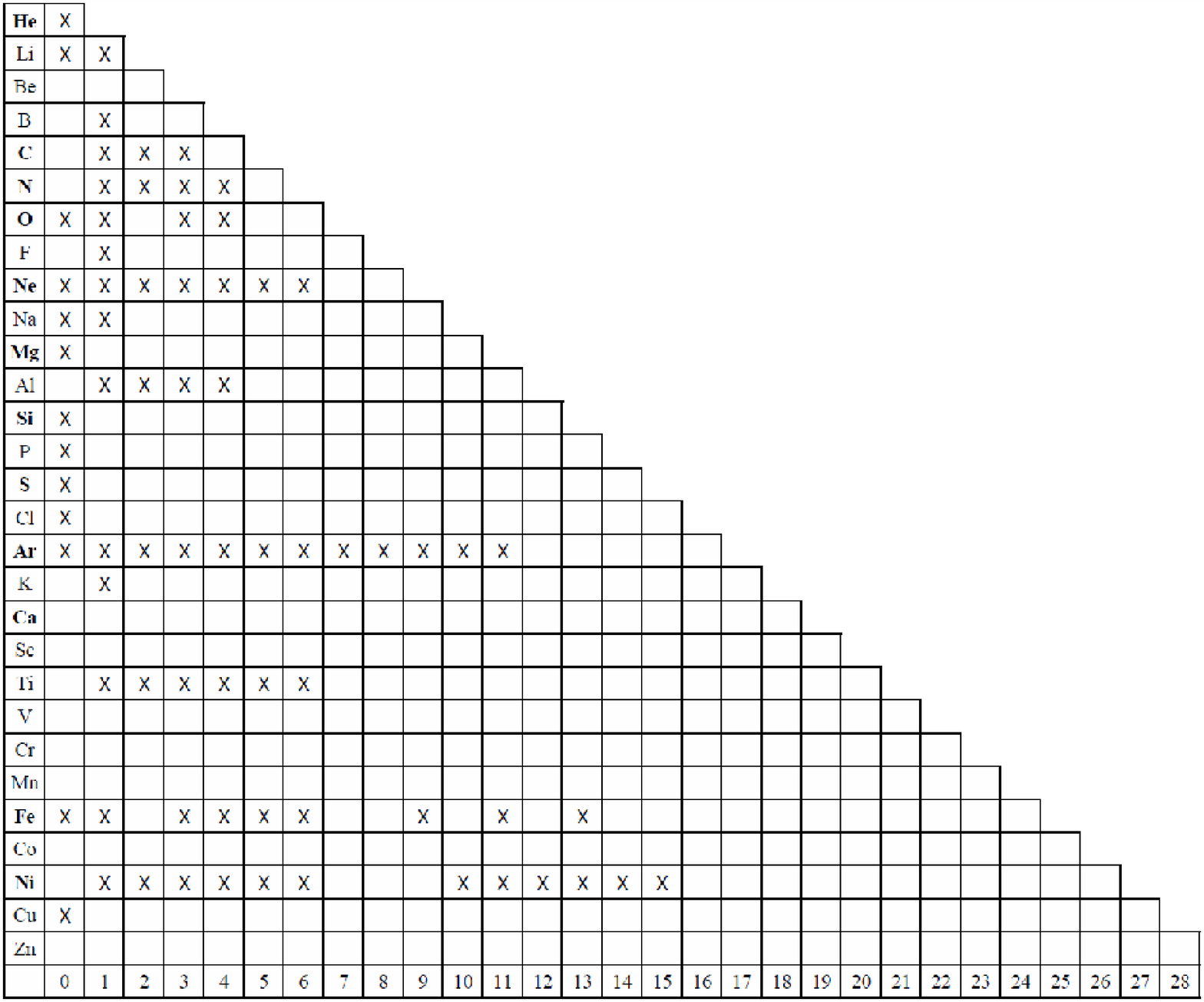}
	\caption{\label{fig:datadouble} Chart summarizing the availability of double ionization data for various ions. The elements are given in rows and each column represents the initial charge state for the reaction, e.g., He$^{0+}$ is represented by the top left corner of the table. The astrophysically abundant elements, He, C, N, O, Ne, Mg, Si, S, Ar, Ca, Fe, and Ni are highlighted in boldface. An `X' in the corresponding box indicates that experimental double ionization data exist for the reaction. Note that isoelectronic sequences fall on the diagonals. The references for these experimental data can be found in Table~\ref{table:refs}. 
}
\end{figure}

\begin{figure}
	\centering \includegraphics[width=0.9\textwidth]{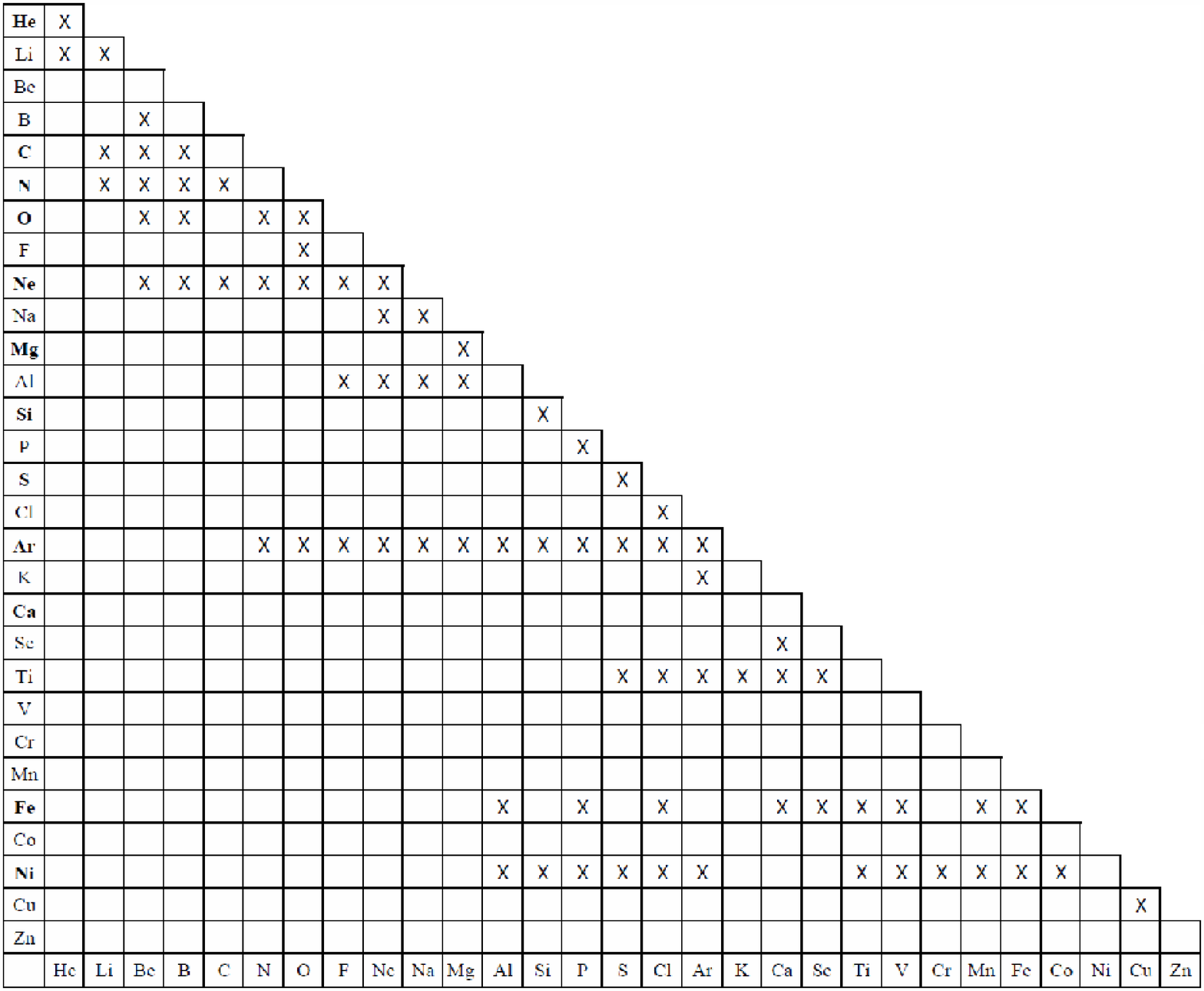}
	\caption{\label{fig:datadoubleiso} Same as Figure~\ref{fig:datadouble}, except that here the columns label isoelectronic sequences. 
	}
\end{figure}

One limitation of the existing data for double and triple ionization is that the measured systems are generally for light and/or low charged-ions. For example, we are not aware of any double ionization measurements for He-like through C-like ions from ions heavier than neon. For M-shell ions, most of the double ionization data come from neutral species or from argon ions. These deficiencies in the double ionization database are illustrated graphically in Figures~\ref{fig:datadouble} and \ref{fig:datadoubleiso}. For triple ionization, we are not aware of any measurements where the initial ion is more than three times ionized. In order to understand how these cross sections evolve along isoelectronic sequences, data are needed for heavier and more highly charged ions. 

The theoretical problem of EIMI is difficult to address because it is a complex many-body problem. However, one advance that could significantly affect the EIMI data would be a revision of the \citet{Kaastra:AAS:1993} IA branching ratios, i.e., the calculation of the probability distribution of the number of ejected electrons following inner shell single ionization. IA is the dominant ionization process for highly charged ions and for high order EIMI. The atomic structure calculations upon which the \citeauthor{Kaastra:AAS:1993} data are based are more than 50 years old. Revising these data should be a tractable theoretical problem. It would also be interesting and useful to calculate similar branching ratios for multiple vacancies and excited states.

\begin{acknowledgments}
This work was supported in part by the NASA Living with a Star Program grant NNX15AB71G and by the NSF Division of Atmospheric and Geospace Sciences SHINE program grant AGS-1459247.
\end{acknowledgments}

\newpage

\begin{deluxetable}{lllll}
	\tabletypesize{\footnotesize}
	\tablecaption{Experimental EIMI data sources. \label{table:refs}}
	\tablewidth{0pt}
	\tablehead{
		\colhead{Order} & 
		\colhead{Sequence} & 
		\colhead{Ion} & 
		\colhead{Reference} & 
		\colhead{Comment} 
	}
	\tablecolumns{5}
	\startdata
	2 		& He 		& He$^{0+}$ 	& \citet{Shah:JPhysB:1988} 	& \nodata \\
	\ditto	& \ditto	& \ditto		& \citet{Rejoub:PRA:2002} 	& \nodata \\
	\ditto 	& \ditto 	& Li$^{1+}$ 	& \citet{Peart:JPhysB:1969} 	& \nodata \\	
	\ditto 	& Li		& Li$^{0+}$ 	& \citet{Jalin:JChemPhys:1973} 	& \nodata \\
	\ditto 	& \ditto	& \ditto		& \citet{Huang:PRA:2002}		& \nodata \\
	\ditto 	& \ditto 	& C$^{3+}$		& \citet{Westermann:PhysScr:1999} & Data from private communication \\
	\ditto 	& \ditto 	& N$^{4+}$		& \citet{Westermann:PhysScr:1999} & Data from private communication \\
	\ditto 	& Be		& B$^{1+}$ 		& \citet{Scheuermann:2001} 	& Reported in \citet{Shevelko:JPhysB:2005} \\
	\ditto 	& \ditto 	& C$^{2+}$		& \citet{Westermann:PhysScr:1999} & Data from private communication \\
	\ditto 	& \ditto 	& N$^{3+}$		& \citet{Westermann:PhysScr:1999} & Data from private communication \\
	\ditto 	& \ditto 	& O$^{4+}$		& \citet{Westermann:PhysScr:1999} & Data from private communication \\
	\ditto 	& \ditto	& Ne$^{6+}$		& \citet{Duponchelle:JPhysB:1997}	& \nodata \\
	\ditto  & B			& C$^{1+}$		& \citet{Zambra:JPhysB:1994} & \nodata \\
	\ditto 	& \ditto	& \ditto		& \citet{Westermann:PhysScr:1999} 	& Data from private communication \\
	\ditto 	& \ditto	& \ditto		& \citet{Lecointre:JPhysB:2013}		& \nodata\\
	\ditto	& \ditto	& N$^{2+}$		& \citet{Westermann:PhysScr:1999}	& Data from private communication \\
	\ditto	& \ditto	& O$^{3+}$		& \citet{Westermann:PhysScr:1999} 	& Data from private communication \\
	\ditto	& \ditto	& Ne$^{5+}$		& \citet{Duponchelle:JPhysB:1997}	& \nodata \\
	\ditto 	& C			& N$^{1+}$		& \citet{Zambra:JPhysB:1994}		& \nodata \\
	\ditto  & \ditto	& \ditto		& \citet{Lecointre:JPhysB:2013} 	& \nodata \\
	\ditto 	& \ditto 	& O$^{2+}$		& \citet{Westermann:PhysScr:1999} & Data from private communication \\
	\ditto 	& \ditto	& Ne$^{4+}$		& \citet{Tinschert:1989} 		& Data from private communication. see also \citet{Shevelko:JPhysB:2005} \\
	\ditto	& N			& O$^{1+}$ 		& \citet{Zambra:JPhysB:1994}	& \nodata \\
	\ditto  & \ditto 	& \ditto 		& \citet{Westermann:PhysScr:1999} & Data from private communication \\
	\ditto  & \ditto 	& \ditto		& \citet{Lecointre:JPhysB:2013}	& \nodata \\
	\ditto	& \ditto	& Ne$^{3+}$		& \citet{Tinschert:1989}		& Data from private communication, see also \citet{Shevelko:JPhysB:2005} \\
	\ditto 	& \ditto	& Ar$^{11+}$	& \citet{Zhang:JPhysB:2002}		& \nodata \\
	\ditto	& O			& O$^{0+}$		& \citet{Ziegler:PSS:1982}		& \nodata \\
	\ditto	& \ditto	& \ditto		& \citet{Thompson:JPhysB:1995}	& \nodata \\
	\ditto	& \ditto	& F$^{1+}$		& \citet{Zambra:JPhysB:1994}	& \nodata \\
	\ditto	& \ditto	& Ne$^{2+}$		& \citet{Tinschert:1989} 		& Data from private communication, see also \citet{Shevelko:JPhysB:2005} \\
	\ditto	& \ditto	&Ar$^{10+}$		& \citet{Zhang:JPhysB:2002}		& \nodata\\
	\ditto	& F			& Ne$^{1+}$		& \citet{Tinschert:1989}		& Data from private communication, see also \citet{Shevelko:JPhysB:2005} \\
	\ditto	& \ditto	& \ditto		& \citet{Zambra:JPhysB:1994}	& \nodata \\
	\ditto 	& \ditto	& Al$^{4+}$		& \citet{Steidl:1999}			& Data from private communication \\
	\ditto	& \ditto	& Ar$^{9+}$		& \citet{Zhang:JPhysB:2002}		& \nodata \\
	\ditto	& Ne		& Ne$^{0+}$		& \citet{Schram:Physica:1966a}	& \nodata \\
	\ditto	& \ditto	& \ditto		& \citet{Krishnakumar:JPhysB:1988}	& \nodata \\
	\ditto	& \ditto	& \ditto		& \citet{Lebius:JPhysB:1989}	& \nodata \\
	\ditto	& \ditto	& \ditto		& \citet{Almeida:JPhysB:1995}	& \nodata \\
	\ditto	& \ditto	& \ditto		& \citet{Rejoub:PRA:2002}		& \nodata \\
	\ditto	& \ditto	& Na$^{1+}$		& \citet{Hirayama:JPSJ:1986}	& See NIFS database at https://dbshino.nifs.ac.jp \\
	\ditto & \ditto 	& Al$^{3+}$		& \citet{Steidl:1999}			& Data from private communication \\ 
	\ditto & \ditto		& Ar$^{8+}$		& \citet{Zhang:JPhysB:2002}		& \nodata \\
	\ditto & Na			& Na$^{0+}$		& \citet{Tate:PR:1934}			& \nodata \\
	\ditto & \ditto		& Al$^{2+}$		& \citet{Steidl:1999}			& Data from private communication \\
	\ditto & \ditto		& Ar$^{7+}$		& \citet{Tinschert:JPhysB:1989}	& \nodata \\
	\ditto & \ditto		& \ditto		& \citet{Rachafi:JPhysB:1991}	& \nodata \\
	\ditto & \ditto		& \ditto		& \citet{Zhang:JPhysB:2002}		& \nodata \\
	\ditto & Mg			& Mg$^{0+}$		& \citet{McCallion:JPhysB:1992}	& \nodata \\
	\ditto & \ditto		& \ditto		& \citet{Boivin:JPhysB:1998}	& \nodata \\
	\ditto & \ditto 	& Al$^{1+}$		& \citet{Steidl:1999}			& Data from private communication \\
	\ditto & \ditto		& Ar$^{6+}$		& \citet{Tinschert:JPhysB:1989} & \nodata \\
	\ditto & \ditto		& \ditto		& \citet{Zhang:JPhysB:2002}		& \nodata \\
	\ditto & Al			& Ar$^{5+}$		& \citet{Tinschert:JPhysB:1989} & \nodata \\
	\ditto & \ditto		& Fe$^{13+}$	& \citet{Hahn:ApJ:2013}			& \nodata \\
	\ditto & \ditto		& Ni$^{15+}$	& \citet{Cherkani:PhysScr:2001} & \nodata \\
	\ditto & Si			& Si$^{0+}$		& \citet{Freund:PRA:1990}		& \nodata \\
	\ditto & \ditto		& Ar$^{4+}$		& \citet{Muller:JPhysB:1985} 	& \nodata \\
	\ditto & \ditto		& Fe$^{12+}$	& \citet{Hahn:ApJ:2011a}		& \nodata \\
	\ditto & \ditto		& Ni$^{14+}$	& \citet{Cherkani:PhysScr:1999} & \nodata \\
	\ditto & P			& P$^{0+}$		& \citet{Freund:PRA:1990}		& \nodata \\
	\ditto & \ditto		& Ar$^{3+}$		& \citet{Muller:PRL:1980}		& \nodata \\
	\ditto & \ditto		& \ditto 		& \citet{Tinschert:JPhysB:1989} & \nodata \\
	\ditto & \ditto		& Fe$^{11+}$	& \citet{Hahn:ApJ:2011}			& \nodata \\
	\ditto & \ditto 	& Ni$^{13+}$	& \citet{Cherkani:PhysScr:2001} & \nodata \\
	\ditto & S			& S$^{0+}$		& \citet{Ziegler:PSS:1982b}		& \nodata \\
	\ditto & \ditto		& \ditto		& \citet{Freund:PRA:1990}		& \nodata \\
	\ditto & \ditto		& Ar$^{2+}$		& \citet{Muller:PRL:1980}		& \nodata \\
	\ditto & \ditto		& \ditto		& \citet{Tinschert:JPhysB:1989}	& \nodata \\
	\ditto & \ditto		& Ti$^{6+}$		& \citet{Hartenfeller:JPhysB:1998} & See also \citet{Shevelko:JPhysB:2006} \\
	\ditto & \ditto		& Ni$^{12+}$	& \citet{Cherkani:PhysScr:1999}	& \nodata \\
	\ditto & Cl			& Cl$^{0+}$ 	& \citet{Freund:PRA:1990}		& \nodata \\
	\ditto & \ditto		& Ar$^{1+}$		& \citet{Muller:PRL:1980}		& \nodata \\
	\ditto & \ditto		& \ditto		& \citet{Muller:JPhysB:1985}	& \nodata \\
	\ditto & \ditto		& \ditto		& \citet{Belic:JPhysB:2010} 	& \nodata \\
	\ditto & \ditto		& Ti$^{5+}$		& \citet{Hartenfeller:JPhysB:1998} & See also \citet{Shevelko:JPhysB:2006} \\
	\ditto & \ditto 	& Fe$^{9+}$		& \citet{Hahn:ApJ:2012}			& \nodata \\
	\ditto & \ditto		& Ni$^{11+}$	& \citet{Cherkani:PhysScr:2001} & \nodata \\
	\ditto & Ar			& Ar$^{0+}$		& \citet{Schram:Physica:1966b}	& \nodata \\
	\ditto & \ditto		& \ditto		& \citet{Krishnakumar:JPhysB:1988} & \nodata \\
	\ditto & \ditto		& \ditto		& \citet{Syage:JPhysB:1991}		& \nodata \\
	\ditto & \ditto 	& \ditto		& \citet{McCallion:JPhysB:1992a} & \nodata \\
	\ditto & \ditto 	& \ditto		& \citet{Straub:PRA:1995}		& \nodata \\
	\ditto & \ditto		& \ditto		& \citet{Rejoub:PRA:2002}		& \nodata \\
	\ditto & \ditto		& K$^{1+}$		& \citet{Hirayama:JPSJ:1986} 	& See NIFS database at https://dbshino.nifs.ac.jp \\
	\ditto & \ditto 	& Ti$^{4+}$ 	& \citet{Hartenfeller:JPhysB:1998} & See also \citet{Shevelko:JPhysB:2006} \\
	\ditto & \ditto		& Ni$^{10+}$	& \citet{Cherkani:PhysScr:1999}	& \nodata \\
	\ditto & K		 	& Ti$^{3+}$ 	& \citet{Hartenfeller:JPhysB:1998} & See also \citet{Shevelko:JPhysB:2006} \\
	\ditto & Ca			& Sc$^{1+}$		& \citet{Jacobi:JPhysB:2005} 	&  \nodata \\
	\ditto & \ditto		& Ti$^{2+}$		& \citet{Hartenfeller:JPhysB:1998} & See also \citet{Shevelko:JPhysB:2006} \\
	\ditto & \ditto		& Fe$^{6+}$		& \citet{Stenke:JPhysB:1999}	& See also \citet{Shevelko:JPhysB:2006} \\
	\ditto & Sc			& Ti$^{1+}$ 	& \citet{Hartenfeller:JPhysB:1998} & See also \citet{Shevelko:JPhysB:2006} \\
	\ditto & \ditto		& Fe$^{5+}$		& \citet{Stenke:JPhysB:1999}	& See also \citet{Shevelko:JPhysB:2006} \\
	\ditto & Ti			& Fe$^{4+}$		& \citet{Stenke:JPhysB:1999}	& See also \citet{Shevelko:JPhysB:2006} \\
	\ditto & \ditto		& Ni$^{6+}$		& \citet{Stenke:NIMB:1995}		& See also \citet{Shevelko:JPhysB:2006} \\
	\ditto & V			& Fe$^{3+}$		& \citet{Stenke:JPhysB:1999}	& See also \citet{Shevelko:JPhysB:2006} \\
	\ditto & \ditto		& Ni$^{5+}$ 	& \citet{Stenke:NIMB:1995}		& See also \citet{Shevelko:JPhysB:2006} \\
	\ditto & Cr			& Ni$^{4+}$ 	& \citet{Stenke:NIMB:1995}		& See also \citet{Shevelko:JPhysB:2006} \\
	\ditto & Mn			& Fe$^{1+}$		& \citet{Stenke:JPhysB:1999}	& See also \citet{Shevelko:JPhysB:2006} \\
	\ditto & \ditto 	& Ni$^{3+}$		& \citet{Stenke:NIMB:1995}		& See also \citet{Shevelko:JPhysB:2006} \\
	\ditto & Fe			& Fe$^{0+}$		& \citet{Shah:JPhysB:1993}		& \nodata \\
	\ditto & \ditto		& Ni$^{2+}$		& \citet{Stenke:NIMB:1995}		& See also \citet{Shevelko:JPhysB:2006} \\
	\ditto & Co			& Ni$^{1+}$		& \citet{Stenke:NIMB:1995}		& See also \citet{Shevelko:JPhysB:2006} \\
	\ditto & Cu			& Cu$^{0+}$		& \citet{Freund:PRA:1990}		& \nodata \\
	\ditto & \ditto 	& \ditto		& \citet{Bolorizadeh:JPhysB:1994}	& \nodata \\
	\ditto & Zn			& \nodata		& \nodata						& \nodata \\
	3		& Li		& Li$^{0+}$		& \citet{Huang:PRL:2003}		& \nodata \\
	\ditto	& B			& C$^{1+}$ 		& \citet{Westermann:PhysScr:1999} & Data from private communication \\
	\ditto 	& \ditto 	& \ditto 		& \citet{Lecointre:JPhysB:2013}	& \nodata \\
	\ditto	& \ditto 	& N$^{2+}$		& \citet{Westermann:PhysScr:1999} & Data from private communication \\
	\ditto 	& \ditto 	& O$^{3+}$		& \citet{Westermann:PhysScr:1999} & Data from private communication \\
	\ditto 	& C			& N$^{1+}$		& \citet{Lecointre:JPhysB:2013} & \nodata \\
	\ditto 	& \ditto 	& O$^{2+}$		& \citet{Westermann:PhysScr:1999} & Data from private communication \\
	\ditto 	& N			& O$^{1+}$		& \citet{Westermann:PhysScr:1999} & Data from private communication \\
	\ditto 	& \ditto 	& \ditto		& \citet{Lecointre:JPhysB:2013}	& \nodata \\
	\ditto	& F			& Ne$^{1+}$		& \citet{Tinschert:1989} 		& Data from private communication \\
	\ditto 	& \ditto	& \ditto		& \nodata						& Unpublished data from E. Salzborn's group \\
	\ditto 	& Ne		& Ne$^{0+}$		& \citet{Schram:Physica:1966a} 	& \nodata \\
	\ditto	& \ditto	& \ditto		& \citet{Krishnakumar:JPhysB:1988}	& \nodata \\
	\ditto	& \ditto	& \ditto		& \citet{Lebius:JPhysB:1989}	& \nodata \\
	\ditto	& \ditto	& \ditto		& \citet{Almeida:JPhysB:1995}	& \nodata \\
	\ditto	& \ditto	& \ditto		& \citet{Rejoub:PRA:2002}		& \nodata \\
	\ditto 	& \ditto 	& Al$^{3+}$		& \citet{Steidl:1999}			& Data from private communication \\
	\ditto 	& Na		& Al$^{2+}$		& \citet{Steidl:1999}			& Data from private communication \\
	\ditto	& Mg		& Mg$^{0+}$		& \citet{McCallion:JPhysB:1992}	& \nodata \\
	\ditto & \ditto		& \ditto		& \citet{Boivin:JPhysB:1998}	& \nodata \\
	\ditto	& \ditto	& Al$^{1+}$		& \citet{Steidl:1999}			& Data from private communication \\
 	\ditto & S			& S$^{0+}$		& \citet{Ziegler:PSS:1982b}		& \nodata \\
 	\ditto & \ditto		& Ar$^{2+}$		& \citet{Muller:PRL:1980} 		& \nodata \\
 	\ditto & \ditto		& Ni$^{12+}$	& \citet{Cherkani:PhysScr:1999}	& \nodata \\
 	\ditto & Cl			& Ar$^{1+}$		& \citet{Muller:PRL:1980}		& \nodata \\
 	\ditto & \ditto 	& \ditto		& \citet{Belic:JPhysB:2010}		& \nodata \\
 	\ditto & Ar			& Ar$^{0+}$		& \citet{Schram:Physica:1966b}	& \nodata \\
 	\ditto & \ditto		& \ditto		& \citet{Krishnakumar:JPhysB:1988} & \nodata \\
 	\ditto & \ditto 	& \ditto		& \citet{Syage:JPhysB:1991}		& \nodata \\
 	\ditto & \ditto		& \ditto		& \citet{McCallion:JPhysB:1992a} & \nodata \\
 	\ditto & \ditto 	& \ditto		& \citet{Almeida:JESRP:1994} 	& \nodata \\
 	\ditto & \ditto 	& \ditto		& \citet{Straub:PRA:1995}		& \nodata \\
 	\ditto & \ditto		& \ditto		& \citet{Rejoub:PRA:2002}		& \nodata \\
 	\ditto & K			& Ti$^{3+}$		& \citet{Hartenfeller:JPhysB:1998} & \nodata \\
	\ditto & Ca			& Sc$^{1+}$		& \citet{Jacobi:JPhysB:2005} 	   & \nodata  \\
 	\ditto & \ditto		& Ti$^{2+}$		& \citet{Hartenfeller:JPhysB:1998} & \nodata \\
 	\ditto & Sc			& Ti$^{1+}$		& \citet{Hartenfeller:JPhysB:1998} & \nodata \\
 	\ditto & Fe			& Fe$^{0+}$		& \citet{Shah:JPhysB:1993}			& \nodata \\
	\ditto & Cu			& Cu$^{0+}$		& \citet{Bolorizadeh:JPhysB:1994}	& \nodata \\
	4		& C			& N$^{1+}$		& \citet{Lecointre:JPhysB:2013}		& \nodata \\
	\ditto	& N			& O$^{1+}$		& \citet{Westermann:PhysScr:1999}	& Data from private communication \\
	\ditto 	& \ditto 	& \ditto		& \citet{Lecointre:JPhysB:2013}		& \nodata \\
	\ditto & Ne			& Ne$^{0+}$		& \citet{Schram:Physica:1966a} 	& \nodata \\
	\ditto	& \ditto	& \ditto		& \citet{Krishnakumar:JPhysB:1988}	& \nodata \\
	\ditto	& \ditto	& \ditto		& \citet{Lebius:JPhysB:1989}	& \nodata \\
	\ditto	& \ditto	& \ditto		& \citet{Almeida:JPhysB:1995}	& \nodata \\
	\ditto	& \ditto	& \ditto		& \citet{Rejoub:PRA:2002}		& \nodata \\
	\ditto & Mg			& Mg$^{0+}$		& \citet{McCallion:JPhysB:1992} & \nodata \\
	\ditto & \ditto 	& Al$^{1+}$		& \citet{Steidl:1999}			& Data from private communication \\
	\ditto & S			& S$^{0+}$		& \citet{Ziegler:PSS:1982b}		& \nodata \\
	\ditto & Cl			& Ar$^{1+}$		& \citet{Muller:PRL:1980}		& \nodata \\
	\ditto & \ditto 	& \ditto		& \citet{Belic:JPhysB:2010}		& \nodata \\
	\ditto & Ca			& Sc$^{1+}$		& \citet{Jacobi:JPhysB:2005} 	&  \nodata \\
	\ditto & Sc			& Ti$^{1+}$		& \citet{Hartenfeller:JPhysB:1998} & \nodata \\
	\ditto & Cu			& Cu$^{0+}$		& \citet{Bolorizadeh:JPhysB:1994}	& \nodata \\
	5		& Ne		& Ne$^{0+}$		& \citet{Schram:Physica:1966a}		& \nodata \\
	\ditto 	& \ditto 	& \ditto		& \citet{Almeida:JPhysB:1995}		& \nodata \\
	\ditto 	& Cl		& Ar$^{1+}$		& \citet{Belic:JPhysB:2010} 		& \nodata \\
	\ditto  & Ar		& Ar$^{0+}$		& \citet{Schram:Physica:1966b}		& \nodata \\
	\ditto	& \ditto	& \ditto		& \citet{McCallion:JPhysB:1992a}	& \nodata \\
	\ditto 	& \ditto	& \ditto		& \citet{Almeida:JESRP:1994}		& \nodata \\
	\ditto  & Ca		& Sc$^{1+}$		& \citet{Jacobi:JPhysB:2005} 		& \nodata \\
	\ditto 	& Cu		& \ditto		& \citet{Bolorizadeh:JPhysB:1994} 	& \nodata \\
	6		& Cl		& Ar$^{1+}$		& \citet{Belic:JPhysB:2010}			& \nodata \\
	\ditto	& Ar		& Ar$^{0+}$		& \citet{Schram:Physica:1966b}		& \nodata \\
	\ditto	& \ditto	& \ditto		& \citet{Almeida:JESRP:1994}		& \nodata \\
	7		& Cl		& Ar$^{1+}$		& \citet{Belic:JPhysB:2010}			& \nodata \\
	\ditto	& Ar		& Ar$^{0+}$		& \citet{Schram:Physica:1966b}		& \nodata \\
	\enddata
	\tablecomments{}
\end{deluxetable}

\clearpage
\newpage

\begin{deluxetable}{L C C C C C C C C C C}
\tabletypesize{\small}
\tablecolumns{10}
\tablewidth{0pc}
\tablecaption{Fitting Formulae for EIMI Cross Sections
\label{table:cross}}
\tablehead{
\colhead{$Z$} &
\colhead{$q_{i}$} &
\colhead{$q_{f}$} &
\colhead{Equation} & 
\colhead{$E_{\mathrm{th}}$} & 
\colhead{$f$} &
\colhead{$p_{0}$} &
\colhead{$p_{1}$} &
\colhead{$p_{2}$} &
\colhead{$p_{3}$} &
\colhead{$p_{4}$} 
}
\startdata
2 & 0 & 2 & 7 & 79.0052 & 1 & 2.6 & 6.501 & 1.743 & -19.656 & 11.647\\
3 & 0 & 2 & 2 & 81.0318 & 1 & 5.8 & 0 & 0 & 0 & 0 \\
3 & 0 & 3 & 1 & 203.486 & 0.00312 & 6.3 & 3 & 1.2 & 1 & 0 \\
3 & 1 & 3 & 7 & 198.094 & 1 & 2.6 & 6.501 & 1.743 & -19.656 & 11.647 \\
4 & 0 & 2 & 2 & 27.5339 & 1 & 1.8 & 0 & 0 & 0 & 0 \\
4 & 0 & 2 & 3 & 123.63  & 0.9999 & 3.6 & 4.5 & 0 & 0 & 0 \\
4 & 1 & 3 & 2 & 172.107 & 1 & 12 & 0 & 0 & 0 & 0 \\
4 & 2 & 4 & 7 & 371.615 & 1 & 2.6 & 6.501 & 1.743 & -19.656 & 11.647
\enddata
\tablecomments{}
\end{deluxetable}

\newpage
\begin{deluxetable}{l l L L L L L L L}
	\tabletypesize{\scriptsize}
	\tablecolumns{9}
	\tablewidth{0pc}
	\tablecaption{Inferred IA Branching Ratios for EIMI of Ar$^{1+}$
		\label{table:far1}}
	\tablehead{
		\colhead{Subshell} &
		\colhead{Source} & 
		\colhead{$\rightarrow$ Ar$^{2+}$} &
		\colhead{$\rightarrow$ Ar$^{3+}$} &
		\colhead{$\rightarrow$ Ar$^{4+}$} & 
		\colhead{$\rightarrow$ Ar$^{5+}$} & 
		\colhead{$\rightarrow$ Ar$^{6+}$} & 
		\colhead{$\rightarrow$ Ar$^{7+}$} &
		\colhead{$\rightarrow$ Ar$^{8+}$} 
	}
	\startdata
	$2p$ & This work					& 0.0004 & 0.7776 & 0.222\phn & 0 & 0 & 0 & 0 \\
	$2p$ & \citet{Kaastra:AAS:1993} 	& 0.0004 & 0.9996 & 0 & 0 & 0 & 0 & 0 \\
	$2s$ & This work					& 0.0004 & 0.0348 & 0.7048 & 0.26  & 0 & 0 & 0 \\
	$2s$ & \citet{Kaastra:AAS:1993} 	& 0.0004 & 0.0348 & 0.9648 & 0 & 0 & 0 & 0
	\enddata
\end{deluxetable}

\newpage

\newpage
\begin{deluxetable}{l l L L L L L L L}
	\tabletypesize{\scriptsize}
	\tablecolumns{10}
	\tablewidth{0pc}
	\tablecaption{Inferred IA Branching Ratios for EIMI of Ar$^{0+}$
		\label{table:far0}}
	\tablehead{
		\colhead{Subshell} &
		\colhead{Source} & 
		\colhead{$\rightarrow$ Ar$^{1+}$} &
		\colhead{$\rightarrow$ Ar$^{2+}$} &
		\colhead{$\rightarrow$ Ar$^{3+}$} & 
		\colhead{$\rightarrow$ Ar$^{4+}$} & 
		\colhead{$\rightarrow$ Ar$^{5+}$} & 
		\colhead{$\rightarrow$ Ar$^{6+}$} &
		\colhead{$\rightarrow$ Ar$^{7+}$} 
	}
	\startdata
	$2p$ & This work					& 0.0003 & 0.7497 & 0.25 & 0 & 0 & 0 & 0 \\
	$2p$ & \citet{Kaastra:AAS:1993} 	& 0.0004 & 0.9997 & 0 & 0 & 0 & 0 & 0 \\
	$2s$ & This work					& 0.0005 & 0.0361 & 0.4293 & 0.46  & 0.038 & 0 & 0 \\
	$2s$ & \citet{Kaastra:AAS:1993} 	& 0.0005 & 0.0361 & 0.9634 & 0 & 0 & 0 & 0 \\
	$1s$ & This work					& 0.0102 & 0.1172 & 0.1001 & 0.5172 & 0.1869 & 0.0544 & 0.014 \\
	$1s$ & \citet{Kaastra:AAS:1993}  & 0.0102 & 0.1172 & 0.1001 & 0.5172 & 0.2009 & 0.0544 & 0
	\enddata
\end{deluxetable}

\newpage

\clearpage
\newpage 
\appendix
\section{Examples}\label{apsec:examples}

Here we provide a more extensive set of examples showing our cross sections compared to EIMI measurements. This set of figures is not comprehensive, as we have only plotted those fits for which we have a tabulation of the experimental data. In some cases, data were not tabulated in the original references. For example, the double ionization data for low charge states of Ti, Fe, and Ni were not available in tabular form and our parameters are based on the fits provided by \citet{Shevelko:JPhysB:2006}. In other such cases, our fits are based on a careful inspection of the figures in the experimental papers, e.g., triple ionization of Ti$^{1+}$, Ti$^{2+}$, and Ti$^{3+}$ as reported by \citet{Hartenfeller:JPhysB:1998}.

In most cases, we include error bars on the experimental data in these figures. However, error bars are suppressed in cases where there are many data points, (e.g., double ionization Ar$^{0+}$) so that the plot would be too cluttered, and whenever the uncertainties are not tabulated in the experimental references. In the latter cases, the experimental papers usually provide a general discussion of the uncertainties in the text. For example, \citet{Freund:PRA:1990} do not tabulate uncertainties, but give a discussion in the text estimating a general uncertainty level of about 10\% on their absolute cross sections.

\clearpage

\bibliography{MEII}

\begin{thebibliography}{98}
\expandafter\ifx\csname natexlab\endcsname\relax\def\natexlab#1{#1}\fi

\bibitem[{Akahori \& Yoshikawa(2010)}]{Akahori:PASJ:2010}
Akahori, T., \& Yoshikawa, K. 2010, Publ.\ Astron.\ Soc.\ Japan, 62, 335

\bibitem[{Almeida {et~al.}(1995)Almeida, Fontes, \&
  Godinho}]{Almeida:JPhysB:1995}
Almeida, D.~P., Fontes, A.~C., \& Godinho, C. F.~L. 1995, J.\ Phys.\ B, 28,
  3335

\bibitem[{Almeida {et~al.}(1994)Almeida, Fontes, Mattos, \&
  Godinho}]{Almeida:JESRP:1994}
Almeida, D.~P., Fontes, A.~C., Mattos, I.~S., \& Godinho, C.~L. 1994, Journal
  of Electron Spectroscopy and Related Phenomena, 67, 503

\bibitem[{Bautista {et~al.}(2003)Bautista, Mendoza, Kallman, \&
  Palmeri}]{Bautista:AA:2003}
Bautista, M.~A., Mendoza, C., Kallman, T.~R., \& Palmeri, P. 2003, A\&A, 403,
  339

\bibitem[{B\'elenger {et~al.}(1997)B\'elenger, Defrance, Salzborn, Shevelko,
  Tawara, \& Uskov}]{Belenger:JPhysB:1997}
B\'elenger, C., Defrance, P., Salzborn, E., Shevelko, V.~P., Tawara, H., \&
  Uskov, D.~B. 1997, J.\ Phys.\ B, 30, 2667

\bibitem[{Belic {et~al.}(1987)Belic, Falk, Timmer, \& Dunn}]{Belic:PRA:1987}
Belic, D.~S., Falk, R.~A., Timmer, C., \& Dunn, G.~H. 1987, Phys.\ Rev.\ A, 36,
  1073

\bibitem[{Belic {et~al.}(2010)Belic, Lecointre, \&
  Defrance}]{Belic:JPhysB:2010}
Belic, D.~S., Lecointre, J., \& Defrance, P. 2010, J.\ Phys.\ B, 43, 185203

\bibitem[{Berakdar(1996)}]{Berakdar:PhysLett:1996}
Berakdar, J. 1996, Phys.\ Lett.\ A, 220, 237

\bibitem[{Boivin \& Srivastava(1998)}]{Boivin:JPhysB:1998}
Boivin, R.~F., \& Srivastava, S.~K. 1998, J.\ Phys.\ B, 31, 2381

\bibitem[{Bolorizadeh {et~al.}(1994)Bolorizadeh, Patton, Shah, \&
  Gilbody}]{Bolorizadeh:JPhysB:1994}
Bolorizadeh, M.~A., Patton, C.~J., Shah, M.~B., \& Gilbody, H.~B. 1994, J.\
  Phys.\ B, 27, 175

\bibitem[{Bradshaw \& Klimchuk(2011)}]{Bradshaw:ApJS:2011}
Bradshaw, S.~J., \& Klimchuk, J.~A. 2011, ApJS, 194, 26

\bibitem[{Bryans {et~al.}(2006)Bryans, Badnell, Gorczyca, Laming, Mitthumsiri,
  \& Savin}]{Bryans:ApJS:2006}
Bryans, P., Badnell, N.~R., Gorczyca, T.~W., Laming, J.~M., Mitthumsiri, W., \&
  Savin, D.~W. 2006, ApJS, 167, 343

\bibitem[{Bryans {et~al.}(2009)Bryans, Landi, \& Savin}]{Bryans:ApJ:2009}
Bryans, P., Landi, E., \& Savin, D.~W. 2009, ApJ, 691, 1540

\bibitem[{Cherkani-Hassani {et~al.}(1999)Cherkani-Hassani, Defrance, \&
  Oualim}]{Cherkani:PhysScr:1999}
Cherkani-Hassani, S., Defrance, P., \& Oualim, E.~M. 1999, Phys.\ Scr., T80,
  292

\bibitem[{Cherkani-Hassani {et~al.}(2001)Cherkani-Hassani, Khouilid, \&
  Defrance}]{Cherkani:PhysScr:2001}
Cherkani-Hassani, S., Khouilid, M., \& Defrance, P. 2001, Phys.\ Scr., T92, 287

\bibitem[{Defrance {et~al.}(2003)Defrance, Kereselidze, \&
  Machavariani}]{Defrance:NIMB:2003}
Defrance, P., Kereselidze, T.~M., \& Machavariani, Z.~S. 2003, Nuclear
  Instruments and Methods in Physics Research B, 205, 405

\bibitem[{Defrance {et~al.}(2000)Defrance, Kereselidze, Machavariani, \&
  Noselidze}]{Defrance:JPhysB:2000}
Defrance, P., Kereselidze, T.~M., Machavariani, Z.~S., \& Noselidze, I.~L.
  2000, J.\ Phys.\ B, 33, 4323

\bibitem[{Duponchelle {et~al.}(1997)Duponchelle, Khouilid, Oualim, Zhang, \&
  Defrance}]{Duponchelle:JPhysB:1997}
Duponchelle, M., Khouilid, M., Oualim, E.~M., Zhang, H., \& Defrance, P. 1997,
  J.\ Phys.\ B, 30, 729

\bibitem[{Fisher {et~al.}(1995)Fisher, Ralchenko, Goldgirsh, Fisher, \&
  Maron}]{Fisher:JPhysB:1995}
Fisher, V., Ralchenko, Y., Goldgirsh, A., Fisher, D., \& Maron, Y. 1995, J.\
  Phys.\ B, 28, 3027

\bibitem[{Freund {et~al.}(1990)Freund, Wetzel, Shul, \&
  Hayes}]{Freund:PRA:1990}
Freund, R.~S., Wetzel, R.~C., Shul, R.~J., \& Hayes, T.~R. 1990, Phys.\ Rev.\
  A, 41, 3575

\bibitem[{Garcia {et~al.}(2009)Garcia, Kallman, Witthoeft, Behar, Mendoza,
  Palmeri, Quinet, Bautista, \& Klapisch}]{Garcia:ApJS:2009}
Garcia, J., {et~al.} 2009, ApJS, 185, 477

\bibitem[{Gorczyca {et~al.}(2006)Gorczyca, Dumitriu, Haso\v{g}lu, Korista,
  Badnell, Savin, \& Manson}]{Gorczyca:ApJ:2006}
Gorczyca, T.~W., Dumitriu, I., Haso\v{g}lu, M.~F., Korista, M.~T., Badnell,
  N.~R., Savin, D.~W., \& Manson, S.~T. 2006, ApJ, 638, 121

\bibitem[{Gorczyca {et~al.}(2003)Gorczyca, Kodituwakku, nad O.~Zatsarinny,
  Badnell, Behar, Chen, \& Savin}]{Gorczyca:ApJ:2003}
Gorczyca, T.~W., Kodituwakku, C.~N., nad O.~Zatsarinny, K. T.~K., Badnell,
  N.~R., Behar, E., Chen, M.~H., \& Savin, D.~W. 2003, ApJ, 592, 636

\bibitem[{G\"otz {et~al.}(2006)G\"otz, Walter, \& Briggs}]{Gotz:JPhysB:2006}
G\"otz, J.~R., Walter, M., \& Briggs, J.~S. 2006, J.\ Phys.\ B, 39, 4365

\bibitem[{Gryzi\'nski(1965)}]{Gryzinksi:PR:1965}
Gryzi\'nski, M. 1965, Phys.\ Rev., 138, 336

\bibitem[{Hahn \& Savin(2015{\natexlab{a}})}]{Hahn:ApJ:2015}
Hahn, M., \& Savin, D.~W. 2015{\natexlab{a}}, ApJ, 800, 68

\bibitem[{Hahn \& Savin(2015{\natexlab{b}})}]{Hahn:ApJ:2015b}
---. 2015{\natexlab{b}}, ApJ, 809, 178

\bibitem[{Hahn {et~al.}(2011{\natexlab{a}})Hahn, Bernhardt, Grieser, Krantz,
  Lestinsky, M\"uller, Novotn\'y, Repnow, Schippers, Wolf, \&
  Savin}]{Hahn:ApJ:2011}
Hahn, M., {et~al.} 2011{\natexlab{a}}, ApJ, 729, 76

\bibitem[{Hahn {et~al.}(2011{\natexlab{b}})Hahn, Grieser, Krantz, Lestinsky,
  M\"uller, Novotn\'y, Repnow, Schippers, Wolf, \& Savin}]{Hahn:ApJ:2011a}
---. 2011{\natexlab{b}}, ApJ, 735, 105

\bibitem[{Hahn {et~al.}(2012)Hahn, Becker, Grieser, Krantz, Lestinsky,
  M\"uller, Novotny, Repnow, Schippers, Spruck, Wolf, \& Savin}]{Hahn:ApJ:2012}
---. 2012, ApJ, 760, 80

\bibitem[{Hahn {et~al.}(2013)Hahn, Becker, Bernhardt, Grieser, Krantz,
  Lestinsky, M\"uller, Novotn\'y, Repnow, Schippers, Spruck, Wolf, \&
  Savin}]{Hahn:ApJ:2013}
---. 2013, ApJ, 767, 47

\bibitem[{Hartenfeller {et~al.}(1998)Hartenfeller, Aichele, Hathiramani,
  Sch\"afer, Steidl, Scheuermann, \& Salzborn}]{Hartenfeller:JPhysB:1998}
Hartenfeller, U., Aichele, K., Hathiramani, D., Sch\"afer, V., Steidl, M.,
  Scheuermann, F., \& Salzborn, E. 1998, J.\ Phys.\ B, 31, 3013

\bibitem[{Haso\v{g}lu {et~al.}(2006)Haso\v{g}lu, Gorczyca, Korista, Manson,
  Badnell, \& Savin}]{Hasoglu:ApJ:2006}
Haso\v{g}lu, M.~F., Gorczyca, T.~W., Korista, K.~T., Manson, S.~T., Badnell,
  N.~R., \& Savin, D.~W. 2006, ApJ, 649, L149

\bibitem[{Hirayama {et~al.}(1986)Hirayama, Oda, Morikawa, Ono, Ikezaki,
  Takayanagi, Wakiya, \& Suzuki}]{Hirayama:JPSJ:1986}
Hirayama, T., Oda, K., Morikawa, Y., Ono, T., Ikezaki, Y., Takayanagi, T.,
  Wakiya, K., \& Suzuki, H. 1986, J.\ Phys.\ Soc.\ Japan, 55, 1411

\bibitem[{Huang {et~al.}(2003)Huang, Wong, Inokuti, Southworth, \&
  Young}]{Huang:PRL:2003}
Huang, M.-T., Wong, W.~W., Inokuti, M., Southworth, S.~H., \& Young, L. 2003,
  Phys.\ Rev.\ Lett., 90, 163201

\bibitem[{Huang {et~al.}(2002)Huang, Zhang, Hasegawa, Southworth, \&
  Young}]{Huang:PRA:2002}
Huang, M.-T., Zhang, L., Hasegawa, S., Southworth, S.~H., \& Young, L. 2002,
  Phys.\ Rev.\ A, 66, 012715

\bibitem[{Jacobi {et~al.}(2005)Jacobi, Knopp, Schippers, Shi, \&
  M\"uller}]{Jacobi:JPhysB:2005}
Jacobi, J., Knopp, H., Schippers, S., Shi, W., \& M\"uller, A. 2005, J.\ Phys.\
  B, 38, 2015

\bibitem[{Jalin {et~al.}(1973)Jalin, Hagermann, \&
  Botter}]{Jalin:JChemPhys:1973}
Jalin, R., Hagermann, R., \& Botter, R. 1973, J.\ Chem.\ Phys., 59, 952

\bibitem[{Jonauskas {et~al.}(2014)Jonauskas, Prancikevi\v{c}ius, \v{S}. Masys,
  \& Kynien\'e}]{Jonauskas:PRA:2014}
Jonauskas, V., Prancikevi\v{c}ius, A., \v{S}. Masys, \& Kynien\'e, A. 2014,
  Phys.\ Rev.\ A, 89, 052714

\bibitem[{Kaastra \& Mewe(1993)}]{Kaastra:AAS:1993}
Kaastra, J.~S., \& Mewe, R. 1993, Astron.\ Astrophys.\ Suppl.\ Ser., 97, 443

\bibitem[{Kim \& Rudd(1994)}]{Kim:PRA:1994}
Kim, Y.-K., \& Rudd, M.~E. 1994, Phys.\ Rev.\ A, 50, 3954

\bibitem[{Kim {et~al.}(2000)Kim, Santos, \& Parente}]{Kim:PRA:2000}
Kim, Y.-K., Santos, J.~P., \& Parente, F. 2000, Phys.\ Rev.\ A, 62, 052710

\bibitem[{Kramida {et~al.}(2016)Kramida, Ralchenko, Reader, \& {NIST ASD Team
  (2016)}}]{NIST:2016}
Kramida, A., Ralchenko, Y., Reader, J., \& {NIST ASD Team (2016)}. 2016, NIST
  Atomic Spectra Database (version 5.4), National Institute of Standards and
  Technology

\bibitem[{Krishnakumar \& Srivastava(1988)}]{Krishnakumar:JPhysB:1988}
Krishnakumar, E., \& Srivastava, S.~K. 1988, J.\ Phys.\ B, 21, 1055

\bibitem[{Ku\v{c}as {et~al.}(2015)Ku\v{c}as, Momkauskaite, \&
  Karazija}]{Kucas:ApJ:2015}
Ku\v{c}as, S., Momkauskaite, A., \& Karazija, R. 2015, ApJ, 810, 26

\bibitem[{Lahmam-Bennani {et~al.}(2010)Lahmam-Bennani, {Staicu Casagrande},
  Naja, {Dal Cappello}, \& Bolognesi}]{Lahmam:JPhysB:2010}
Lahmam-Bennani, A., {Staicu Casagrande}, E.~M., Naja, A., {Dal Cappello}, C.,
  \& Bolognesi, P. 2010, J.\ Phys.\ B, 43, 105201

\bibitem[{Lebius {et~al.}(1989)Lebius, Binder, Koslowski, Wiesemann, \&
  Huber}]{Lebius:JPhysB:1989}
Lebius, H., Binder, J., Koslowski, H.~R., Wiesemann, K., \& Huber, B.~A. 1989,
  J.\ Phys.\ B, 22, 83

\bibitem[{Lecointre {et~al.}(2013)Lecointre, Kouzakov, Belic, Defrance, Popov,
  \& Shevelko}]{Lecointre:JPhysB:2013}
Lecointre, J., Kouzakov, K.~A., Belic, D.~S., Defrance, P., Popov, Y.~V., \&
  Shevelko, V.~P. 2013, J.\ Phys.\ B, 46, 205201

\bibitem[{Lotz(1969)}]{Lotz:ZPhys:1969}
Lotz, W. 1969, Z.\ Physik, 220, 466

\bibitem[{Magee {et~al.}(1995)}]{Magee:ASP:1995}
Magee, N.~H., {et~al.} 1995, in ASP Conference Series, Vol.~78, Astrophysical
  Applications of Powerful New Databases, ed. S.~J. Adelman \& W.~L. Wiese (San
  Francisco: ASP), 51, http://aphysics2.lanl.gov/tempweb/

\bibitem[{McCallion {et~al.}(1992{\natexlab{a}})McCallion, Shah, \&
  Gilbody}]{McCallion:JPhysB:1992a}
McCallion, P., Shah, M.~B., \& Gilbody, H.~B. 1992{\natexlab{a}}, J.\ Phys.\ B,
  25, 1061

\bibitem[{McCallion {et~al.}(1992{\natexlab{b}})McCallion, Shah, \&
  Gilbody}]{McCallion:JPhysB:1992}
---. 1992{\natexlab{b}}, J.\ Phys.\ B, 25, 1051

\bibitem[{Mendoza {et~al.}(2004)Mendoza, Kallman, Bautista, \&
  Palmeri}]{Mendoza:AA:2004}
Mendoza, C., Kallman, T.~R., Bautista, M.~A., \& Palmeri, P. 2004, A\&A, 414,
  377

\bibitem[{M\"uller(1986)}]{Muller:PhysLett:1986}
M\"uller, A. 1986, Phys.\ Lett., 113A, 415

\bibitem[{M\"uller(2005)}]{Muller:NIMB:2005}
---. 2005, Nuclear Instruments and Methods in Physics Research B, 233, 141

\bibitem[{M\"{u}ller(2008)}]{Muller:Book:2008}
M\"{u}ller, A. 2008, in Advances in Atomic, Molecular, and Optical Physics, 55,
  ed. E.~Arimondo, P.~Berman, \& C.~Lin (London: Elsevier), 293

\bibitem[{M\"uller \& Frodl(1980)}]{Muller:PRL:1980}
M\"uller, A., \& Frodl, R. 1980, Phys.\ Rev.\ Lett., 44, 29

\bibitem[{M\"uller {et~al.}(1985)M\"uller, Tinschert, Achenbach, Becker, \&
  Salzborn}]{Muller:JPhysB:1985}
M\"uller, A., Tinschert, K., Achenbach, C., Becker, R., \& Salzborn, E. 1985,
  J.\ Phys.\ B, 18, 3011

\bibitem[{M\"uller {et~al.}(1988)M\"uller, Tinschert, Hofmann, Salzborn, \&
  Dunn}]{Muller:PRL:1988}
M\"uller, A., Tinschert, K., Hofmann, G., Salzborn, E., \& Dunn, G.~H. 1988,
  Phys.\ Rev.\ Lett., 61, 70

\bibitem[{M\"uller {et~al.}(2017)M\"uller, Bernhardt, {Borovik, Jr.}, Buhr,
  Hellhund, Holste, Kilcoyne, Klumpp, Martins, Ricz, Seltmann, Viefhaus, \&
  Schippers}]{Muller:ApJ:2017}
M\"uller, A., {et~al.} 2017, ApJ, 836, 166

\bibitem[{Palmeri {et~al.}(2003)Palmeri, Mendoza, Kallman, Bautista, \&
  Mel\'endez}]{Palmeri:AA:2003}
Palmeri, P., Mendoza, C., Kallman, T.~R., Bautista, M.~A., \& Mel\'endez, M.
  2003, A\&A, 410, 359

\bibitem[{Palmeri {et~al.}(2011)Palmeri, Quinet, Mendoza, Bautista, Garc\'ia,
  Witthoeft, \& Kallman}]{Palmeri:AA:2011}
Palmeri, P., Quinet, P., Mendoza, C., Bautista, M.~A., Garc\'ia, J., Witthoeft,
  M.~C., \& Kallman, T.~R. 2011, A\&A, 525, 59

\bibitem[{Palmeri {et~al.}(2012)Palmeri, Quinet, Mendoza, Bautista, Garc\'ia,
  Witthoeft, \& Kallman}]{Palmeri:AA:2012}
---. 2012, A\&A, 543, 44

\bibitem[{Patnaude {et~al.}(2009)Patnaude, Ellison, \&
  Slane}]{Patnaude:ApJ:2009}
Patnaude, D.~J., Ellison, D.~C., \& Slane, P. 2009, ApJ, 696, 1956

\bibitem[{Peart \& Dolder(1969)}]{Peart:JPhysB:1969}
Peart, B., \& Dolder, K.~T. 1969, J.\ Phys.\ B, 2, 1169

\bibitem[{Pindzola {et~al.}(2010)Pindzola, Ballance, Robicheaux, \&
  Colgan}]{Pindzola:JPhysB:2010}
Pindzola, M.~S., Ballance, C.~P., Robicheaux, F., \& Colgan, J. 2010, J.\
  Phys.\ B, 43, 105204

\bibitem[{Pindzola \& Loch(2017)}]{Pindzola:JPhysB:2017}
Pindzola, M.~S., \& Loch, S.~D. 2017, J.\ Phys.\ B, 50, 085203

\bibitem[{Pindzola {et~al.}(2011)Pindzola, Ludlow, Ballance, Robicheaux, \&
  Colgan}]{Pindzola:JPhysB:2011}
Pindzola, M.~S., Ludlow, J.~A., Ballance, C.~P., Robicheaux, F., \& Colgan, J.
  2011, J.\ Phys.\ B, 44, 105202

\bibitem[{Pindzola {et~al.}(2009)Pindzola, Ludlow, Robicheaux, Colgan, \&
  Griffin}]{Pindzola:JPhysB:2009}
Pindzola, M.~S., Ludlow, J.~A., Robicheaux, F., Colgan, J., \& Griffin, D.~C.
  2009, J.\ Phys.\ B, 42, 215204

\bibitem[{Rachafi {et~al.}(1991)Rachafi, Belic, Duponchelle, Jureta, Zambra,
  Hui, \& Defrance}]{Rachafi:JPhysB:1991}
Rachafi, S., Belic, D.~S., Duponchelle, M., Jureta, J., Zambra, M., Hui, Z., \&
  Defrance, P. 1991, J.\ Phys.\ B, 24, 1037

\bibitem[{Reale \& Orlando(2008)}]{Reale:ApJ:2008}
Reale, F., \& Orlando, S. 2008, ApJ, 684, 715

\bibitem[{Rejoub {et~al.}(2002)Rejoub, Lindsay, \& Stebbings}]{Rejoub:PRA:2002}
Rejoub, R., Lindsay, B.~G., \& Stebbings, R.~F. 2002, Phys.\ Rev.\ A, 65,
  042713

\bibitem[{Scheuermann {et~al.}(2001)Scheuermann, Jacobi, Salzborn, \&
  M\"uller}]{Scheuermann:2001}
Scheuermann, F., Jacobi, J., Salzborn, E., \& M\"uller, A. 2001, unpublished

\bibitem[{Schram(1966)}]{Schram:Physica:1966b}
Schram, B.~L. 1966, Physica, 32, 197

\bibitem[{Schram {et~al.}(1966)Schram, Boerboom, \&
  Kistemaker}]{Schram:Physica:1966a}
Schram, B.~L., Boerboom, A. J.~H., \& Kistemaker, J. 1966, Physica, 32, 185

\bibitem[{Shah {et~al.}(1988)Shah, Elliot, McCallion, \&
  Gilbody}]{Shah:JPhysB:1988}
Shah, M.~B., Elliot, D.~S., McCallion, P., \& Gilbody, H.~B. 1988, J.\ Phys.\
  B, 21, 2751

\bibitem[{Shah {et~al.}(1993)Shah, McCallion, Okuno, \&
  Gilbody}]{Shah:JPhysB:1993}
Shah, M.~B., McCallion, P., Okuno, K., \& Gilbody, H.~B. 1993, J.\ Phys.\ B,
  26, 2393

\bibitem[{Shevelko \& Tawara(1995{\natexlab{a}})}]{Shevelko:PhysScr:1995}
Shevelko, V.~P., \& Tawara, H. 1995{\natexlab{a}}, Phys.\ Scr., 52, 649

\bibitem[{Shevelko \& Tawara(1995{\natexlab{b}})}]{Shevelko:JPhysB:1995}
---. 1995{\natexlab{b}}, J.\ Phys.\ B, 28, L589

\bibitem[{Shevelko {et~al.}(2005)Shevelko, Tawara, Scheuermann, Fabian,
  M\"uller, \& Salzborn}]{Shevelko:JPhysB:2005}
Shevelko, V.~P., Tawara, H., Scheuermann, F., Fabian, B., M\"uller, A., \&
  Salzborn, E. 2005, J.\ Phys.\ B, 38, 525

\bibitem[{Shevelko {et~al.}(2006)Shevelko, Tawara, Tolstikhina, Scheuermann,
  Fabian, M\"uller, \& Salzborn}]{Shevelko:JPhysB:2006}
Shevelko, V.~P., Tawara, H., Tolstikhina, I.~Y., Scheuermann, F., Fabian, B.,
  M\"uller, A., \& Salzborn, E. 2006, J.\ Phys.\ B, 39, 1499

\bibitem[{Steidl {et~al.}(1999)Steidl, Aichele, Hartenfeller, Hathiramani,
  Pindzola, Scheuermann, Westermann, \& Salzborn}]{Steidl:1999}
Steidl, M., Aichele, K., Hartenfeller, U., Hathiramani, D., Pindzola, M.~S.,
  Scheuermann, F., Westermann, M., \& Salzborn, E. 1999, in Abstract Book of
  21st Int.\ Conf.\ on the Physics of Photonic, Electronic, and Atomic
  Collisions, ed. Y.~Itikawa, K.~Okuno, H.~Tanaka, A.~Yagishita, \&
  M.~Matsuzawa (Sendai Japan), 362

\bibitem[{Stenke {et~al.}(1999)Stenke, Hartenfeller, Aichele, Hathiramani,
  Steidl, \& Salzborn}]{Stenke:JPhysB:1999}
Stenke, M., Hartenfeller, U., Aichele, K., Hathiramani, D., Steidl, M., \&
  Salzborn, E. 1999, J.\ Phys.\ B, 32, 3641

\bibitem[{Stenke {et~al.}(1995)Stenke, Hathiramani, Hofmann, Shevelko, Steidl,
  V\"olpel, \& Salzborn}]{Stenke:NIMB:1995}
Stenke, M., Hathiramani, D., Hofmann, G., Shevelko, V.~P., Steidl, M.,
  V\"olpel, R., \& Salzborn, E. 1995, Nucl.\ Inst.\ Meth.\ Phys.\ Res.\ B, 98,
  138

\bibitem[{Stolte {et~al.}(2016)Stolte, Jonauskas, Lindle, {Sant'Anna}, \&
  Savin}]{Stolte:ApJ:2016}
Stolte, W.~C., Jonauskas, V., Lindle, D.~W., {Sant'Anna}, M.~M., \& Savin,
  D.~W. 2016, ApJ, 818, 149

\bibitem[{Straub {et~al.}(1995)Straub, Renault, Lindsay, Smith, \&
  Stebbings}]{Straub:PRA:1995}
Straub, H.~C., Renault, P., Lindsay, B.~G., Smith, K.~A., \& Stebbings, R.~F.
  1995, Phys.\ Rev.\ A, 52, 1115

\bibitem[{Syage(1991)}]{Syage:JPhysB:1991}
Syage, J.~A. 1991, J.\ Phys.\ B, 24, 527

\bibitem[{Talukder {et~al.}(2009)Talukder, Haque, \&
  Uddin}]{Talukder:EurPhys:2009}
Talukder, M.~R., Haque, A. K.~F., \& Uddin, M.~A. 2009, Eur.\ Phys.\ J.\ D, 53,
  133

\bibitem[{Tate \& Smith(1934)}]{Tate:PR:1934}
Tate, J.~T., \& Smith, P.~T. 1934, Phys.\ Rev., 46, 773

\bibitem[{Thompson {et~al.}(1995)Thompson, Shah, \&
  Gilbody}]{Thompson:JPhysB:1995}
Thompson, W.~R., Shah, M.~B., \& Gilbody, H.~B. 1995, J.\ Phys.\ B, 28, 1321

\bibitem[{Tinschert(1989)}]{Tinschert:1989}
Tinschert, K. 1989, PhD thesis, Justus-Liebig Universit\"at Giessen

\bibitem[{Tinschert {et~al.}(1989)Tinschert, M\"uller, Phaneuf, Hofmann, \&
  Salzborn}]{Tinschert:JPhysB:1989}
Tinschert, K., M\"uller, A., Phaneuf, R.~A., Hofmann, G., \& Salzborn, E. 1989,
  J.\ Phys.\ B, 22, 1241

\bibitem[{Westermann {et~al.}(1999)Westermann, Scheuermann, Aichele,
  Hartenfeller, Hathiramani, Steidl, \& Salzborn}]{Westermann:PhysScr:1999}
Westermann, M., Scheuermann, F., Aichele, K., Hartenfeller, U., Hathiramani,
  D., Steidl, M., \& Salzborn, E. 1999, Physica Scripta, T80, 285

\bibitem[{Yu {et~al.}(1992)Yu, Rachafi, Jureta, \& Defrance}]{Yu:JPhysB:1992}
Yu, D.~J., Rachafi, S., Jureta, J., \& Defrance, P. 1992, J.\ Phys.\ B, 25,
  4593

\bibitem[{Zambra {et~al.}(1994)Zambra, Belic, Defrance, \&
  Yu}]{Zambra:JPhysB:1994}
Zambra, M., Belic, D., Defrance, P., \& Yu, D.~J. 1994, J.\ Phys.\ B, 27, 2383

\bibitem[{Zhang {et~al.}(2002)Zhang, Cherkani-Hassani, B\'elenger, Duponchelle,
  Khouilid, Oualim, \& Defrance}]{Zhang:JPhysB:2002}
Zhang, H., Cherkani-Hassani, S., B\'elenger, C., Duponchelle, M., Khouilid, M.,
  Oualim, E.~M., \& Defrance, P. 2002, J.\ Phys.\ B, 35, 3829

\bibitem[{Ziegler {et~al.}(1982{\natexlab{a}})Ziegler, Newman, Goeller, Smith,
  \& Stebbings}]{Ziegler:PSS:1982b}
Ziegler, D.~L., Newman, J.~H., Goeller, L.~N., Smith, K.~A., \& Stebbings,
  R.~F. 1982{\natexlab{a}}, Planet.\ Space Sci., 30, 1269

\bibitem[{Ziegler {et~al.}(1982{\natexlab{b}})Ziegler, Newman, Smith, \&
  Stebbings}]{Ziegler:PSS:1982}
Ziegler, D.~L., Newman, J.~H., Smith, K.~A., \& Stebbings, R.~F.
  1982{\natexlab{b}}, Planet.\ Space Sci., 30, 451

\end{thebibliography}

\end{document}